\documentclass[aps,prc,reprint,superscriptaddress,nofootinbib]{revtex4-2}

\usepackage{graphicx} 
\usepackage{xcolor}
\usepackage[normalem]{ulem}
\usepackage{hyperref} 
\hypersetup{colorlinks=true, citecolor=blue, urlcolor=blue, linkcolor=blue}
\usepackage{mathtools}
\usepackage{amssymb}

\begin{document}

\newcommand{\deriv}[2]{\frac{d #1}{d #2}}
\newcommand{\pderiv}[2]{\frac{\partial #1}{\partial #2}}
\newcommand{\ket}[1]{\left| #1 \right\rangle}
\newcommand{\bra}[1]{\left\langle #1 \right|}
\newcommand{\braket}[2]{\left\langle #1 \big| #2 \right\rangle}
\newcommand{\ketbra}[2]{\left| \left. #1 \right\rangle \left\langle #2 \right. \right|}

\newcommand{\elem}[3]{^{#1}_{#2}\mathrm{#3}}
\newcommand{\units}[1]{\ \mathrm{[#1]}}
\newcommand{\unitsn}[1]{\ \mathrm{#1}}
\newcommand{\bv}[1]{\mathbf{#1}}
\newcommand{\sfrac}[2]{{\textstyle\frac{#1}{#2}}}
\newcommand{\xten}[1]{\times 10^{#1}}

\newcommand{\ah}[0]{{\hat{a}}}
\newcommand{\ahd}[0]{{\hat{a}^\dagger}}
\newcommand{\ch}[0]{{\hat{c}}}
\newcommand{\chd}[0]{{\hat{c}^\dagger}}
\newcommand{\bh}[0]{{\hat{\beta}}}
\newcommand{\bhd}[0]{{\hat{\beta}^\dagger}}


\title{An exact time-dependent generator coordinate method with projection for spontaneous fission studies}


\author{N.-W. T. Lau}
\email[]{ngee\_wein.lau@l2it.in2p3.fr}
\affiliation{Universit\'e de Toulouse, CNRS/IN2P3, L2IT, Toulouse, France}

\author{G. Scamps}
\affiliation{Universit\'e de Toulouse, CNRS/IN2P3, L2IT, Toulouse, France}

\author{P. Tan}
\affiliation{CEA, DES, IRESNE, DER, SPRC, LEPh, 13115 Saint-Paul-l\`es-Durance, France}

\author{R. N. Bernard}
\affiliation{CEA, DES, IRESNE, DER, SPRC, LEPh, 13115 Saint-Paul-l\`es-Durance, France}


\date{\today}

\begin{abstract}
Microscopic descriptions of fission dynamics through the barrier rely on different theoretical frameworks depending on the fission regime, with the time-dependent generator coordinate method (TDGCM) commonly used for induced fission and semiclassical approaches such as the Wentzel-Kramers-Brillouin (WKB) method widely employed for spontaneous fission. In this work, we develop a fully microscopic approach to spontaneous fission based on an exact formulation of the TDGCM, avoiding both the Gaussian overlap approximation and semiclassical treatments. By combining the exact TDGCM with new projection tools and a quasistatic description of spontaneous fission, we derive spontaneous fission half-lives from a one-dimensional potential energy surface. A detailed analysis of the model parameters shows that they do not introduce uncontrolled phenomenological effects. Remaining discrepancies with experimental half-lives are related to possible limitations of the underlying microscopic interaction and to the choice of basis states in the generator coordinate method. The new framework provides a microscopic description of spontaneous fission with reduced empirical input and offers a pathway toward improved predictions through future developments of nuclear interactions, collective coordinates, and the handling of many-body dynamics.
\end{abstract}


\maketitle

\section{\label{sec:intro}Introduction}

Ever since its discovery in the 1930s, the process of nuclear fission has drawn a great deal of interest from scientific communities across the world seeking to characterise it with experimental observations, explain it with theoretical models, and develop new applications of its capabilities to produce vast amounts of energy and exotic isotopes. However, in the modern era, rapid and continuous development of theoretical simulations is needed in order to keep pace with the demands of the experimental sector. A particular subdomain where models are in serious need of improvement is the study of spontaneous fission. While theoretical calculations of spontaneous fission lifetimes of nuclei are continuously becoming more accurate, the vast majority of models contain phenomenological parameters which are, to various degrees, fitted to existing data. This significantly limits their usefulness when attempting to simulate the spontaneous fission of exotic nuclei that cannot be produced and observed in laboratory conditions; such nuclei are of great importance for active research topics such as the search for new superheavy elements and the understanding of stellar nucleosynthesis. As a result, new descriptions of spontaneous fission based on microscopic models with minimal phenomenological input are in high demand \cite{bender2020}. \\

In a nucleus undergoing spontaneous fission, the potential energy barrier keeping the nucleons in a bound state is low enough for the nucleus to pass through it via the quantum mechanical phenomenon of tunnelling. However, the conventional means of determining the spontaneous fission lifetime in models relies on the Wentzel-Kramers-Brillouin (WKB) method \cite{merzbacher1998quantum,schunck2016} to obtain a semiclassical approximation for the penetration probability based on the action of a path on the collective potential energy surface (PES) crossing the fission barrier. This action is defined in turn from the ``collective inertia'' of the nuclear system and the quantum mechanical ``zero-point energy'' corrections \cite{sadhukhan2013,baran2015,sadhukhan2020}, which are themselves often calculated approximately. The spontaneous fission lifetime is exponentially sensitive to changes in these quantities, as well as the interaction used to generate the PES; thus, predictions by different models for the same nucleus \cite{giuliani2013,schunck2022} often vary immensely. An alternate approach which is fully quantum mechanical and not reliant on the semiclassical approximations will hence be less susceptible to these subjective influences. Furthermore, the semiclassical method requires the energy of the system to be very close to its ground state, which makes it unsuitable for sub-barrier fission of excited nuclei. \\

One of the authors previously proposed a method which considered the spontaneously fissioning nucleus as a ``quasistatic'' (or metastable) state in a complex absorbing potential \cite{scamps2015}; the imaginary contribution to the state's energy was then used to determine its resonance lifetime as a direct measure of the spontaneous fission lifetime. However, the requirement for an exact solution of the time-dependent Schr\"odinger equation limited the initial application of this model to artificial local potentials in one and two dimensions of collective nuclear shape coordinates. In the present work, this ``quasistatic approach'' is reformulated and extended to be compatible with a one-dimensional microscopic PES calculated with Hartree-Fock-Bogoliubov (HFB) theory. In order to simulate the propagation of the fissioning nucleus across the PES, a new formalism of exact TDGCM developed by the lead author \cite{lauphdthesis} is employed. The smoothness of the input PES (referring to its absence of discontinuities, see Refs. \cite{dubray2012,lau2022,zdeb2021}), required for the exact TDGCM, is guaranteed by using an innovative method proposed recently by Carpentier \textit{et al.} \cite{carpentier2024}, which enforces continuity by directly constraining the overlaps between adjacent solutions on the path. Finally, new mathematical techniques are introduced to project arbitrary GCM wavefunctions in a manner that restores the generator coordinates, normally treated only via their expectation values, as well-defined observables. This leads to a robust method to modify the nonlocal Hamiltonian kernel and add the complex absorbing potential needed for the quasistatic approach to spontaneous fission. \\

Section \ref{sec:methodology} introduces the necessary theory for the current work: the formalism of the TDGCM, the concept of the quasistatic approach, and the techniques of projection onto the generator coordinates. Section \ref{sec:applications} then explains the assembly of these different components into a coherent model that describes the process of spontaneous fission and extracts the spontaneous fission lifetime of the nucleus. The results of the model are analysed in detail in Section \ref{sec:results} to ensure that they are reasonable and numerically sound, followed by the discussion of some more complex issues requiring further investigation and a comparison of the predicted values to experimental data. Finally, Section \ref{sec:conclusion} summarises the presented work and proposes avenues for future research.

\section{\label{sec:methodology}Methodology}

\subsection{\label{sec:methodology:tdgcm}The time-dependent generator coordinate method}

The generator coordinate method (GCM) was developed in the 1950s by Hill, Wheeler and Griffin \cite{hill1953,griffin1957} to describe collective motion in nuclei, while its time-dependent equivalent (TDGCM) was first applied to nuclear fission by Berger \textit{et al.} in 1984 \cite{berger1984}. As such, the theory is widely known and comprehensive treatments are found in textbooks and review articles \cite{ringandschuck,verriere2020}. The brief summary of the TDGCM that follows will serve to highlight the key components for this work, and to define the quantities and notation which will be used. \\

The first step of the TDGCM is the formulation of an ansatz representing the system of interest, written as
\begin{align}
\ket{\Psi(t)} = \int_{\mathcal{Q}} dq\ f(q,t) \ket{\phi(q)}\!, \label{eq:tdgcm_ansatz}
\end{align}
where $q \in \mathcal{Q}$ is a vector of one or more generator coordinates which parametrises the generator states $\ket{\phi(q)}$\footnote{In the context of nuclear fission, the generator states are typically static; only the weight function $f(q,t)$ varies with time. While the theory of TDGCM where both the weight function and basis states depend on time dates back to 1983 \cite{reinhard1983}, practical applications to nuclear fission such as \cite{Li2023} are extremely difficult.}. These states are normalised but not necessarily orthogonal, such that $\big|\langle \phi(q) | \phi(q)\rangle\big| = 1$ but in general $0~\leq~\big|\langle \phi(q) | \phi(q') \rangle\big|~\leq~1$. The complex-valued ``weight function'' $f(q,t)$ quantifies the ansatz as a mixture of generator states at any given time, and it is bound by the normalisation condition imposed on the ansatz:
\begin{align}
1 & = \braket{\Psi(t)}{\Psi(t)} \notag \\
& = \iint_{\mathcal{Q}} dq\,dq'\ f(q,t) f^*(q',t) \braket{\phi(q\vphantom')}{\phi(q')}\!.
\end{align}

Within the context of nuclear fission, the coordinates $q$ are the degrees of freedom chosen to constrain the collective shape of the compound nucleus, while the generator states $\ket{\phi(q)}$ are mean-field solutions for the nuclear wavefunction calculated across a mesh of constraints on $q$. These solutions are commonly visualised with a potential energy surface (PES), which portrays the variation of the energy $E$ of the nucleus as a function of the constrained coordinates $q$ according to $E(q) = \bra{\phi(q)} \hat{H} \ket{\phi(q)}$. The majority of TDGCM-based fission descriptions to date have been performed with two generator coordinates $q \equiv (q_{20},q_{30})$, constraining the axial quadrupole and octupole moments of the nuclear density. These collective degrees of freedom govern the elongation and mass asymmetry of the nucleus respectively\footnote{It is increasingly common to use the dimensionless variables $\beta_2,\beta_3$ instead of the multipole moments, but the two sets of coordinates are usually linearly dependent.}. This choice allows the TDGCM to simulate multiple fission paths, distinguishing between symmetric and asymmetric fission and providing the probability distribution to obtain different primary fission fragments. \\

Applying a variational principle to the ansatz in Eq.~\eqref{eq:tdgcm_ansatz} expresses the evolution of $f(q,t)$ in terms of the Hill-Wheeler equation
\begin{align}
\int_{\mathcal{Q}}\! dq' \bigg[H(q,q') - i\hbar N(q,q') \deriv{}{t}\bigg] f(q',t) = 0, \label{eq:hill_wheeler}
\end{align}
which holds for any value of $q \in \mathcal{Q}$, and introduces the Hamiltonian kernel $H(q,q')$ and overlap kernel $N(q,q')$ respectively as
\begin{align}
H(q,q') & = \bra{\phi(q\vphantom')} \hat{H} \ket{\phi(q')}\! \text{ and} \label{eq:hamiltonian_kernel}\\
N(q,q') & = \braket{\phi(q\vphantom')}{\phi(q')}\!. \label{eq:overlap_kernel}
\end{align}
These two quantities are typically calculated as symmetric real $N\times N$ matrices, where $N$ is the number of elements in $\mathcal{Q}$. The ``exact'' solution of the Hill-Wheeler equation, at its most general, requires the overlap kernel to be positive definite such that it is diagonalisable with strictly positive eigenvalues; this admits the definition of its square root $N^{1/2}(a,q)$ and inverse square root $N^{-1/2}(q,a)$. A ``natural basis'' can then be constructed with the states
\begin{align}
\ket{\psi(a)} & = \int_{\mathcal{Q}}\! dq\ N^{-1/2}(q,a) \ket{\phi(q)}\!, \label{eq:natural_basis}
\end{align}
while the weight function in the natural basis is
\begin{align}
g(a,t) & = \int_{\mathcal{Q}}\! dq\ N^{1/2}(a,q) f(q,t). \label{eq:natural_basis_weight_function}
\end{align}
Applying $N^{-1/2}$ to the left of Eq. \eqref{eq:hill_wheeler} and utilising the above two definitions simplifies the Hill-Wheeler equation into the form
\begin{align}
\int da'\ H_C(a,a') g(a',t) & = i\hbar \deriv{}{t} g(a,t), \label{eq:coll_schrodinger}
\end{align}
introducing the collective Hamiltonian
\begin{align}
H_C(a,a') & = \iint_{\mathcal{Q}^2}\! dq\, dq'\, N^{-1/2}(a,q) H(q,q') N^{-1/2}(q',a') \notag \\
& = \bra{\psi(a\vphantom')} \hat{H} \ket{\psi(a')}\!. \label{eq:coll_hamiltonian}
\end{align}
The natural basis states $\ket{\psi(a)}$ form an orthonormal basis of the space spanned by the set of generator states $\ket{\phi(q)}$. It should be noted that the coordinates $a$ that parametrise the natural basis states, unlike the generator coordinates $q$, do not have any significance regarding the collective shape of the nuclear system; a natural basis state $\ket{\psi(a)}$ is in general a linear combination of multiple generator states $\ket{\phi(q)}$ corresponding to different collective configurations. \\

The time evolution of the system is encoded in the weight function $g(a,t)$, which is determined by solving the equation of motion given in Eq. \eqref{eq:coll_schrodinger}. The inverse of the transformation in Eq. \eqref{eq:natural_basis_weight_function} then yields the dynamics of the original weight function $f(q,t)$. The collective Schr\"odinger equation is conventionally solved with iterative numerical methods such as the Crank-Nicholson method \cite{crank1947}, but such techniques are limited to short periods of time evolution. On the other hand, an exact solution can be obtained by treating the equation as a matrix-vector system of ordinary differential equations of the form
\begin{align}
H_C \cdot \bv{g}(t) & = i\hbar \deriv{\bv{g}}{t}, \label{eq:coll_schrodinger2}
\end{align}
the solution of which is given by separation of variables as
\begin{align}
\bv{g}(t) & = e^{-i H_C t/\hbar} \cdot \bv{g}(0).
\end{align}
This approach is useful under the condition that the collective Hamiltonian $H_C \equiv H_C(a,a')$ is diagonalisable, which allows its matrix exponential to be calculated analytically. \\

\subsection{\label{sec:methodology:quasistatic}Quasistatic approach to spontaneous fission}

The existence of the exact solution to the TDGCM enables its application to spontaneous fission via an adaptation of the quasistatic approach proposed in \cite{scamps2015}, which is a fully microscopic and quantum-mechanical way to obtain the spontaneous fission lifetime of a nucleus. This method defines a modification to the Hamiltonian given by
\begin{align}
\hat{H}_q = \hat{H} + \hat{V}_q + i\hat{W}, \label{eq:quasistatic_hamiltonian}
\end{align}
where $\hat{V}_q$ and $\hat{W}$ are real operators. $\hat{V}_q$ is chosen to ``flatten'' the potential outside the fission barrier(s) of the nucleus; if this is not done, the sharply decreasing energy in the saddle-to-scission and post-scission regions can erroneously suppress the evolution of the wavefunction because its propagating waveform must then be strongly excited with a high frequency that may not be representable in a discrete basis. $i\hat{W}$ is defined to introduce a negative imaginary ``absorption'' potential in the same region to simulate an open boundary condition, allowing the spontaneously fissioning wavefunction to exit the system without reflecting off the edges of the domain \cite{nakatsukasa2005,reinhard2006}. The time-independent solutions of this system, obtained by diagonalising the Hamiltonian to obtain its eigenstates $\ket{\varphi^q_i}$\footnote{Since the modified Hamiltonian is non-Hermitian, its left and right eigenstates $\langle\varphi_i^{qL}|$ and $|\varphi_i^{qR}\rangle$ are not complex conjugates of each other. However, the necessary change in formalism is straightforward will be neglected for the sake of simplicity.}, are then the quasistatic or metastable states with corresponding complex energies $E'_i = E_i - \frac{i}{2}\Gamma_i$. \\

Each quasistatic state is said to be in resonance with an eigenstate of the bound system which has the same real energy $E_i$. These bound states  $\ket{\varphi^c_i}$ are the static solutions of a different modified Hamiltonian $\hat{H}_c = \hat{H} + \hat{V}_c$, where $\hat{V}_c$ is once again real, and is chosen to extrapolate an infinite quadratic potential around the ground state well of the nuclear PES. By comparing the real energies of the two sets of eigenstates $\ket{\varphi^q_i}$ and $\ket{\varphi^c_j}$ and furthermore verifying that the overlap $\langle\varphi^q_i|\varphi^c_j\rangle \approx 1$, the quasistatic state in resonance with the ground state of the nucleus can be identified. The width of this resonance is given by $\Gamma_i$, and the lifetime of the resonance follows straightforwardly as $\tau_i = \hbar/\Gamma_i$. Since the definition of the imaginary potential $i\hat{W}$ simulates the wavefunction exiting the system via spontaneous fission, $\tau_i$ can then be interpreted as the spontaneous fission lifetime of the nucleus. This is typically expressed in terms of the spontaneous fission half-life $(t_{1/2})_i = \tau_i \ln 2$, and this convention will be subsequently upheld to facilitate comparison with other studies. \\

In Ref.~\cite{scamps2015}, the newly proposed quasistatic approach was applied to ``toy model'' potentials in one and two dimensions constructed to simulate fission barriers. In this work, the method is adapted for use with microscopically-generated one-dimensional PESs within the exact GCM formalism. However, a significant obstacle is encountered when applying the method to this new context: the quantity of interest to be modified to generate the quasistatic and bound states is not the Hamiltonian $\hat{H}$, but instead its kernel $H(q,q')$. Since the correcting potentials $V_q(q)$, $W(q)$ and $V_c(q)$ are defined as functions of the generator coordinates, the process of constructing equivalent matrices $V_q(q,q')$, etc. to modify the Hamiltonian kernel is undefined. A rigorous solution to this problem is obtained via newly formulated projection techniques which are presented below.

\subsection{\label{sec:methodology:projection}Projection onto the generator coordinates}

Within the context of nuclear physics theory, projection is typically used to restore symmetries broken by the mean-field model used to describe the nucleus. Recently, projection has been employed to determine nontrivial many-body observables, such as the probability of nucleon transfer \cite{simenel2010} or the spin of the fission fragments \cite{bertsch2019}. In the GCM, a notable category of operators that has been overlooked until now as a projection candidate is that of the generator coordinates themselves. When chosen from a microscopic PES, the generator states $\ket{\phi(q)}$ are parametrised by generator coordinates $q$: in reality, these are only constraints applied to the expectation value of the corresponding operator $\hat{Q}$ such that $\langle\hat{Q}\rangle \equiv \langle \phi(q)|\hat{Q}|\phi(q)\rangle = q$. The standard deviation of this quantity can be calculated by computing $\langle \hat{Q}^2 \rangle$, but no other information is known about the true distribution of $\ket{\phi(q)}$ in the generator coordinate space. Despite this uncertainty, the generator coordinates of a state are often used to infer properties of the nucleus it represents such as whether or not it is scissioned and the identities of its fission fragments if so; as a result, addressing this deficiency of information with projection is an important task. \\

A projection operator $\hat{P}$ (or ``projector'') is defined for a one-body operator $\hat{Q}$ and a particular eigenvalue of that operator $Q$ (chosen from a continuous real spectrum) by the integral
\begin{align}
\hat{P}_{\hat{Q}}(Q) & = \frac{1}{2\pi} \int_{-\infty}^\infty d\varphi\ e^{i\varphi(\hat{Q} - Q)}.
\end{align}
This expression bears a strong resemblance to the Fourier transform of the Dirac delta function $\delta(x - x_0)$, except that the argument of the function contains an operator. Presuming the intuitive form of $\hat{Q}$ as
\begin{align}
\hat{Q} & \equiv \int dQ\ Q \big|\psi(Q)\big\rangle\big\langle\psi(Q)\big|
\end{align} 
where the integral extends over all eigenvalues $Q$ corresponding to eigenstates $\ket{\psi(Q)}$, the action of the projector on an arbitrary wavefunction $\ket{\Psi}$ is given by $\hat{P}_{\hat{Q}}(Q) \ket{\Psi} = n_q \ket{\psi(Q)}$, where the ``occupancy'' $n_q = \langle\psi(Q)|\Psi\rangle$ is a number. The projector $\hat{P}_{\hat{Q}}(Q)$ therefore ``picks out'' the component of $\ket{\Psi}$ which is parallel to the eigenstate $\ket{\psi(Q)}$ corresponding to the eigenvalue $Q$. \\

In this work, the projector for the quadrupole moment operator of the nucleon density $\hat{Q}_{20}$ is employed. In contrast with commonly projected quantities (angular momentum, parity, particle number), $\hat{Q}_{20}$ has an infinite and continuous spectrum of eigenvalues $q_{20}$. (For ease of notation, the reader should henceforth assume that $\hat{Q} \equiv \hat{Q}_{20}$ and $Q \equiv Q_{20}$ unless otherwise specified.) To determine and apply the projector $\hat{P}_{\hat{Q}}(Q)$ in a computational context, a bounded and discrete equivalent must be constructed.

\subsubsection{\label{sec:methodology:discrete_projection}Definition of a discrete projection operator}

Discretisation of the operator $\hat{Q}$ is performed by restricting its eigenvalues to occupy points on an infinite mesh with a constant real spacing $\Delta Q > 0$ between points, such that its eigenstates can be labelled by an integer index $\{\ket{\psi_k}\}_{k\in\mathbb{Z}}$ with corresponding eigenvalues $k\Delta Q$. The action of the operator on an eigenstate is then given by $\hat{Q} \ket{\psi_k} = k\Delta Q \ket{\psi_k}$. \\

For an observable operator with a discrete eigenbasis, the appropriate projector can be expressed using the geometric series representation of the Kronecker delta function
\begin{align}
\hat{P}_{\hat{Q}}(Q) & = \lim_{N\rightarrow\infty} \frac{1}{N} \sum_{j=1}^N e^{2i\pi j(\hat{Q} - Q)/N\Delta Q},
\end{align}
where the assumption that the value of $(\hat{Q} - Q)/\Delta Q$ is an integer after the projector acts on a state follows from the discretisation. Since the sum becomes infinite when the limit $N\rightarrow\infty$ is taken, $N$ must be fixed to a finite value for a numerical implementation. It can be shown that this introduces periodicity into the projector, so that the ``finite'' projector is equal to an infinite sum of the complete projectors which is periodic over some length $L\Delta Q$:
\begin{align}
\hat{P}^L_{\hat{Q}}(Q) & = \frac{1}{L} \sum_{j=1}^L e^{2i\pi j(\hat{Q} - Q)/L\Delta Q} \label{eq:projector_discrete_bounded} \\
& = \sum_{n=-\infty}^\infty\!\!\! \hat{P}_{\hat{Q}}(Q + nL\Delta Q). \notag
\end{align}
However, suppose furthermore that the set of eigenvalues $k\Delta Q$ is bounded according to $k_\mathrm{min} \leq k \leq k_\mathrm{max}$ for some integers $k_\mathrm{min}$ and $k_\mathrm{max}$. As long as $L > k_\mathrm{max} - k_\mathrm{min}$, the finite projector $\hat{P}^L_{\hat{Q}}(Q)$ will be fully equivalent to the complete projector $\hat{P}_{\hat{Q}}(Q)$ within this domain. Since the number of generator states is necessarily finite, it is entirely reasonable to truncate the eigenbasis of $\hat{Q}$ correspondingly, and hence a finite projector $\hat{P}^L_{\hat{Q}}(Q)$ that is sufficient for calculations can always be found. \\

The choices of $\Delta Q$ and $L$ are made by the user, which unavoidably introduces subjective input into the process. In particular, a smaller $\Delta Q$ yields a better approximation of the original continuous eigenvalue spectrum with the discrete mesh. However, for a given observable operator $\hat{Q}$ with particular bounds in its eigenvalues, the size of $L$ necessary to avoid the aforementioned periodicity issues is inversely proportional to that of $\Delta Q$. Thus, $\Delta Q$ should be no smaller than required, because decreasing it increases the number of terms to be summed over in $\hat{P}^L_{\hat{Q}}(Q)$.

\subsubsection{Matrix expression of the axial quadrupole moment operator}

To apply the projector $\hat{P}^L_{\hat{Q}}(Q)$ to a state $\ket{\Psi}$, both the operator $\hat{Q}$ and the state should be expressed in the same basis. Furthermore, since the operator $\hat{Q}$ appears within a matrix exponential in the projector as shown in Eq. \eqref{eq:projector_discrete_bounded}, it must be diagonalised to produce an analytic result. The operator of the axial quadrupole moment is given in the three-dimensional Cartesian basis as $\hat{Q}_{20} = C_{20}(2\hat{z}^2 - (\hat{x}^2 + \hat{y}^2))$, where $C_{20}$ is a constant coefficient which varies by convention; however, the generator states $\ket{\phi(q)}$ in this work are produced by the code \textsc{HFBaxial} \cite{hfbaxial} in an axially-symmetric harmonic oscillator (HO) basis with a finite number of principal shells $N_\mathrm{sh}$. The transformation of $\hat{Q}_{20}$ is thus specific to the cylindrical HO basis defined by \textsc{HFBaxial}. The final matrix expression of $\hat{Q}_{20}$ and some remarks on its derivation can be found in Appendix~\ref{sec:appendix_q20}.

\subsubsection{Application to generator states}

Given an arbitrary state $\ket{\Psi}$, a discretised and bounded operator $\hat{Q}$, and its projector $\hat{P}^L_{\hat{Q}}(Q)$, the probability of measuring the state by applying $\hat{Q}$ and obtaining the outcome $Q$ is simply 
\begin{align}
\mathrm{Pr}(Q) & = \bra{\Psi} \hat{P}^L_{\hat{Q}}(Q) \ket{\Psi}\!.
\end{align}
Using the definition of $\ket{\Psi}$ according to the TDGCM given in Eq. \eqref{eq:tdgcm_ansatz} gives
\begin{align}
\mathrm{Pr}(Q) & = \iint_{\mathcal{Q}^2} dq\,dq'\ f^*(q,t) f(q',t) P(Q;\,q,q'), \label{eq:projected_probability}
\end{align}
introducing the key quantity $P(Q;\,q,q')$ henceforth known as the ``probability kernel'':
\begin{align}
P(Q;\,q,q') & = \bra{\phi(q)\vphantom'} \hat{P}^L_{\hat{Q}}(Q) \ket{\phi(q')}\!.
\end{align}
Inserting the definition of the projector according to Eq. \eqref{eq:projector_discrete_bounded} yields
\begin{align}
P(Q;\,q,q') & = \frac{1}{L} \sum_{j=1}^L \bra{\phi(q\vphantom')} e^{2i\pi j(\hat{Q} - Q)/L\Delta Q} \ket{\phi(q')}\!.
\end{align}
For practical computations, it is useful to consider the equivalent form
\begin{align}
P(Q;\,q,q') & = \mu\! \sum_{j=-M}^M\! R^M_j(q,q') e^{-2i\pi \mu jQ/\Delta Q}, \label{eq:probability_kernel}
\end{align}
shifting the limits of the sum with the symmetries of the complex exponential such that $2M+1 = L$, introducing the shorthand $\mu = 1/(2M+1)$, and defining the ``rotation kernel'' $R^M_j(q,q')$ and rotation operator $\hat{R}^M_j$ as
\begin{align}
R^M_j(q,q') & = \bra{\phi(q)\vphantom'} \hat{R}^M_j \ket{\phi(q')}\! \text{ and} \label{eq:rotation_kernel} \\
\hat{R}^M_j & = e^{2i\pi \mu j\hat{Q}/\Delta Q} 
\end{align}
respectively. In order to evaluate the matrix exponential in the rotation operator, it is assumed that the matrix expression of the quadrupole moment operator $\hat{Q}$ is diagonalisable according to $\hat{Q} = ABA^\dagger$ (where $A$ and $B$ are unitary and diagonal matrices respectively), resulting in a practical form of the rotation operator
\begin{align}
\hat{R}^M_j & = A e^{2i\pi \mu jB/\Delta Q} A^\dagger. \label{eq:rotation_operator}
\end{align}
This two-layered definition avoids redundant calculations of the rotation kernels. Both $R_j^M(q,q')$ and $P(Q;q,q')$ are vectors of $N\times N$ complex Hermitian matrices with $N(N+1)/2$ unique elements, indexed by $j$ and $Q$ respectively. The shift of the sum and the use of $R_{-j}^M(q,q') = [R_j^M(q,q')]^*$ means the entire process requires the calculation of $M+1$ matrices for the rotation kernels, followed by $2M+1$ matrices for the probability kernels. \\

The rotation kernels are computed using the Pfaffian formalism presented by Bertsch and Robledo in \cite{bertsch2012}, with some simplications proposed more recently by Carlsson and Rotureau \cite{carlsson2021}. The Pfaffian expression for the operator kernel used in this work can be found in Appendix~\ref{sec:appendix_pfaffian}. \\

A noteworthy advantage of this new formulation of projection compared to existing projection after variation (PAV) techniques is the ability to project mixtures of mean-field states $\ket{\Psi}$ onto a desired basis. While the calculation of the probability kernels $P(Q;q,q')$ as defined in Eq. \eqref{eq:probability_kernel} is a sizeable computational task, doing so permits the ``measurement'' via projection of arbitrary superpositions of generator states, which is much more suitable for beyond mean-field methods such as GCM and TDGCM. By contrast, previous PAV methods effectively calculate only the diagonal components $P(Q;q,q)$ of the probability kernels, and can thus only compute the projections of pure mean-field states $\ket{\phi(q)}$.


\section{\label{sec:applications}Applications}

\subsection{\label{sec:applications:1d_fission_path}Calculation of the one-dimensional fission path}

The most crucial ingredient for a TDGCM-based model of nuclear fission is a PES calculated for the nuclide under investigation, as the generator states $\ket{\phi(q)}$ are chosen from its solution wavefunctions. However, a particular requirement for TDGCM is that this underlying PES must be continuous, formally expressed by the condition
\begin{align}
\lim_{q' \rightarrow q} N(q,q') = \lim_{q' \rightarrow q} \braket{\phi(q)}{\phi(q')} = 1
\end{align}
for any $q,q'$ within the domain of the PES. Violation of this requirement gives rise to artefacts known as ``discontinuities'', which are especially problematic for TDGCM because they act as impenetrable barriers that the evolving wavefunction cannot cross. More detailed analysis of discontinuities can be found in Refs. \cite{dubray2012,lau2022,zdeb2021}. However, a new approach to generate discontinuity-free PESs in one dimension was proposed by Carpentier \textit{et al.} \cite{carpentier2024} in the form of the ``link'' and ``drop'' methods, which generate continuous 1D fission paths by imposing constraints on the overlaps of neighbouring states. A particular advantage of this approach is its ability to avoid the complex ``scission line'' discontinuity which occurs on conventional PESs, extending the fission path until the primary fragments are well separated. \\

\begin{figure}[h!]
\includegraphics[width=8.6cm]{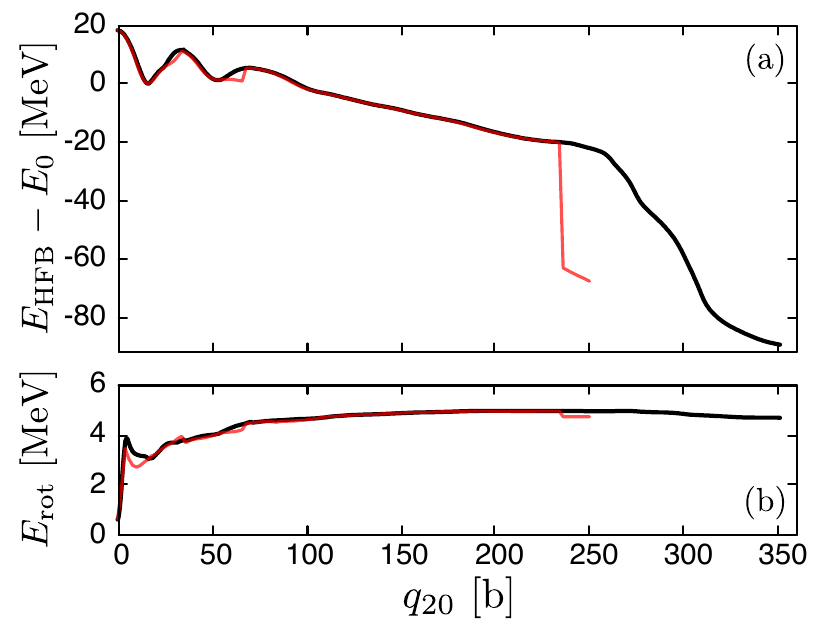}
\caption{\label{fig:256Fm_pes} (a) The one-dimensional fission path (PES) of $\elem{256}{}{Fm}$. The red line indicates the initial PES which was generated by constraining $q_{20}$ from 0 b to 250 b in 2 b increments. The black line is the continuous PES subsequently produced using the methods explained in the main text. Panel (b) depicts the \textit{a posteriori} rotational energy correction.}
\end{figure}

The principal nuclide investigated was $\elem{256}{}{Fm}$, whose 1D PES calculated with the Gogny D1S interaction is shown in Figure \ref{fig:256Fm_pes}(a). For all \textsc{HFBaxial} calculations, the maximum number of shells in the HO basis was set to $N_\mathrm{sh}=16$, with oscillator lengths fixed to $b_\perp=2.1$~fm and $b_z=3.0$~fm. The initial fission path, drawn with a red line, was constructed from HFB solutions for the nucleus with constraints on $q_{20}$ from 0 to 250~b in 2~b increments, where the each solution was used as the initial state for the next calculation. A combination of the link and drop methods, with an overlap constraint of 0.99, was then used to construct the continuous PES drawn in black. The link method has some subjective input because it connects an initial state to a destination state chosen by the user. Therefore the drop method, which does not require a destination but instead follows decreasing energy gradients, was preferred whenever possible. The starting points for this technique were the energy maxima of the original PES: the spherical solution at $q_{20} = 0$~b, and the states at the top of the first and second fission barriers. This produced several segments of a continuous PES which were connected using the link method. The sharp drop appearing in the original PES around $q_{20} \sim 240$~b is the scission line discontinuity is typical to most PESs; as desired, the drop method is able to maintain a continuous path that proceeds through scission up to a much larger elongation. The rotational correction $E_\mathrm{rot}$ to the HFB energy, shown in Figure \ref{fig:256Fm_pes}(b), was determined from the perturbative Yoccoz moment of inertia \cite[\S11.4.4]{ringandschuck} from the resulting wavefunctions and hence was not considered in the variational process. These moments of inertia were multiplied by a phenomenological factor of 1.4 to compensate for the method's known tendency to underestimate the true values calculated with angular momentum projection \cite{warda2002,rodriguezguzman2014}. The resulting changes to the fission barrier are accounted for by incorporating $E_\mathrm{rot}$ into the quasistatic and confining potentials in Section \ref{sec:applications:potentials}. \\

The very high overlaps between the states of the PES produced in this way, however, present a different problem for the TDGCM. The requirement for the overlap kernel $N(q,q')$ to be positive definite is mathematically satisfied as long as none of the generator states are linearly dependent. In practice, however, choosing a set of generator states with overlaps between them approaching 1 leads to the appearance of very small eigenvalues which cause numerical instabilities. This is conventionally addressed in static GCM calculations by truncating eigenvectors corresponding to eigenvalues smaller than some threshold parameter $\epsilon$ from the natural basis \cite{bonche1990,robledo2019,martinezlarraz2022}. Unfortunately, doing this sacrifices the exact invertibility of the natural basis transformation described by $N^{1/2}(a,q)$ and $N^{-1/2}(q,a)$ as mentioned in Ref. \cite[pp.~72--73]{ballyphd}. Given that this property is critical to ensure the good behaviour of the time-evolving GCM state, this work favours instead choosing a sparser selection of HFB states from the PES as the set of generator states. This lowers the overlaps between the states, thus increasing the size of the smallest eigenvalues and rendering the truncation of the eigenbasis unnecessary. \\

The full effects of varying the density of the set of generator states (subsequently referred to as ``mesh density'') on TDGCM and GCM calculations are investigated by comparing and analysing results obtained from sets of generator states with a range of different mesh densities. These sets are chosen from the PES shown in Figure~\ref{fig:256Fm_pes} using a simple ``pick every $M$th point'' method, with $M$ ranging from 3 to 8; increasing $M$ results in a sparser set of generator states representing the same overall PES with lower overlaps between adjacent states, as detailed in Table~\ref{tab:mesh_densities}. \\

\begin{table}[h!]
\begin{tabular}{c|c|c}
Step size $M$ & No. states $N$ & Overlap $\bar{N}$ $\pm$ $\sigma$ \\
\hline
[1] & [495] & [$0.990$] \\
3 & 165 & $0.914 \pm 0.007$ \\
4 & 124 & $0.853 \pm 0.010$ \\
5 & 99  & $0.781 \pm 0.011$ \\
6 & 83  & $0.702 \pm 0.020$ \\
7 & 71  & $0.620 \pm 0.033$ \\
8 & 62  & $0.537 \pm 0.035$
\end{tabular}
\caption{\label{tab:mesh_densities}Properties of the different selections of sets of generator states from the PES calculated for $\elem{256}{}{Fm}$. The full PES with $M=1$ was not used in calculations due to its extremely high density. $\bar{N}$ is the average overlap between neighbouring states across the entire set, and is followed by its standard deviation.}
\end{table}

The ``overlap'' $\bar{N}$ specified in Table~\ref{tab:mesh_densities} is the average of the overlaps between neighbouring states across the entire set. This value is subsequently used to differentiate the different data sets, as it is directly linked to the degree of linear independence between the generator states (henceforth, the term ``overlap'' will refer to this quantity unless stated otherwise). Each row of the table (excluding the first) represents a different set of input data used in subsequent calculations to test the effects of varying the size of the set of generator states. \\

The Hamiltonian and overlap kernels defined in Eqs. \eqref{eq:hamiltonian_kernel} and \eqref{eq:overlap_kernel} were calculated for each data set using a modified version of a program provided by L. Robledo alongside \textsc{HFBaxial} \cite{hfbaxial} which determines the kernels using the generalised Wick's theorem \cite{robledo1994} and the mixed-density formalism \cite{robledo2010}. For examples of other publications using the original version of this tool, see \cite{rodriguezguzman2012,rodriguezguzman2023}.

\subsection{Projection onto $Q_{20}$}

In order to apply the quasistatic approach outlined in Section \ref{sec:methodology:quasistatic} to these data sets, it is first necessary to calculate the probability kernels $P(Q; q,q')$ as defined in Eq. \eqref{eq:probability_kernel}, which are themselves obtained as sums over the rotation kernels $R_j^L(q,q')$ given in Eq. \eqref{eq:rotation_kernel}, for each data set. These calculations are the most computationally intensive part of the overall method. \\

\begin{figure}[h!]
\includegraphics[width=8.6cm]{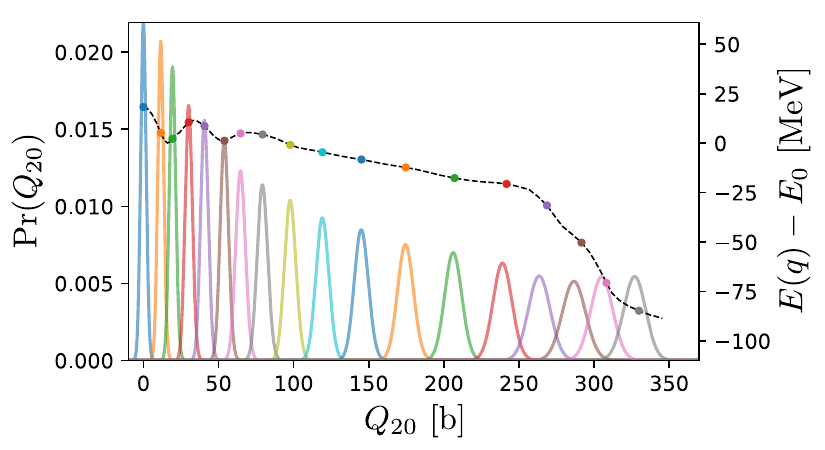}
\caption{\label{fig:hfb_states_proj_256Fm_7s}The distribution of $Q_{20}$ for various HFB wavefunctions from the PES, which are drawn with translucent coloured lines. To avoid excess clutter, the data set with $\bar{N} = 0.62$ was used, and only every $4^{\mathrm{th}}$ state is shown. The black dashed line indicates the shape of the PES (with energy according to the right $y$-axis), and the coloured circles along the PES mark the constrained expectation values $q_{20} \equiv \langle\hat{Q}_{20}\rangle$ for the plotted state with the corresponding colour.}
\end{figure}

Even before putting the assembled projection kernels to use in dynamic calculations, they can be employed to visualise the shape of each generator state's probability distribution in $Q_{20}$. It is intuitively expected that each state from the PES $\ket{\phi(q)}$, parametrised by the constrained expectation value $q$ of $\hat{Q}_{20}$, has a distribution in the shape of a Gaussian centred on $q$, with a width $\sigma$ that depends on the state's location in the PES. These projected probability distributions are shown for a selection of HFB wavefunctions in Figure \ref{fig:hfb_states_proj_256Fm_7s}. This graph depicts the probabilities for every $4^{\mathrm{th}}$ HFB state in the $\bar{N} = 0.62$ data set, which is sufficient to illustrate the Gaussian nature of the projected distributions. The constrained expectation value $q_{20}$, which is depicted using circles of corresponding colours along the black dashed line of the PES, matches the peak of each Gaussian except at very large deformations where a small systematic offset becomes apparent. More detailed investigation has revealed that this error decreases with the number of principal shells $N_\mathrm{sh}$ in the harmonic oscillator basis; with $N_\mathrm{sh}=16$ in this work, the magnitude of the deviation is expected to have minimal effects on the final results. \\

\subsection{\label{sec:applications:potentials}Definition of the quasistatic and confining potential}

The quasistatic approach to describing spontaneous fission requires specific modifications to the Hamiltonian as described in Eq. \eqref{eq:quasistatic_hamiltonian}; in the original context of the time-independent Schr\"odinger equation in \cite{scamps2015}, these modifications were applied with a local correcting function $V'(q)$. When working with the TDGCM, this translates to the addition of a term $V'(q,q')$ to the Hamiltonian kernel $H(q,q')$ in a way that preserves its important properties. However, the definition of a suitable matrix $V'(q,q')$ from a local function $V'(q)$ is not straightforward because there is no inherent principle to define the off-diagonal terms of the matrix. The projection techniques introduced in Section \ref{sec:methodology:projection}, however, provide a natural and rigorous framework for this process. \\

First, the correcting function is redefined as $V'(Q)$ to depend on the ``true'' collective coordinate(s) after projection, rather than the unprojected generator coordinate $q$. An appropriate ``correcting kernel'' can then be defined using the projection operator defined in Eq. \eqref{eq:projector_discrete_bounded}:
\begin{align}
V'(q,q') & = \bra{\phi(q\vphantom')} \Big(\sum_Q \hat{P}^L_{\hat{Q}}(Q) V'(Q)\Big) \ket{\phi(q')}\!, \label{eq:correcting_kernel}
\end{align}
where the sum over $Q$ iterates over the discretised and bounded eigenvalues of $\hat{Q}$. Employing the definition of the probability kernel given in Eq. \eqref{eq:probability_kernel} leads to the convenient form
\begin{align}
V'(q,q') & = \sum_Q P(Q;\, q,q') V'(Q).
\end{align}
The function $V'(Q)$ is then chosen following the original proposition of the quasistatic approach as
\begin{align}
V'(Q) & = V_q(Q) + iW(Q) - E_\mathrm{rot}(Q),
\end{align}
where $V_q(Q)$ and $W(Q)$ are real continuous functions that describe the flattening and absorption potentials, respectively, and $E_\mathrm{rot}(Q)$ is the \textit{a posteriori} rotational energy correction introduced in Section \ref{sec:applications:1d_fission_path} and plotted in Figure \ref{fig:256Fm_pes}(b), linearly interpolated to produce a function defined for all values of $Q$. \\

The real potential $V_q(Q)$ used to ``flatten'' the post-barrier potential of the compound nucleus is chosen in a programmatic fashion depending on the PES in question. To allow quasistatic states which populate the post-barrier region, the ``flattening gap'' parameter $\Delta$ is introduced; flattening is applied for all $Q > q_\Delta$, where $q_\Delta$ is the generator coordinate of the least deformed generator state to have energy $E(q_\Delta) \equiv H(q_\Delta,q_\Delta)$ below the ground state energy according to $E(q_\Delta) < E_0 - \Delta$. In order to maintain the height of the overall potential after this point at $E_0 - \Delta$, the flattening potential is defined by
\begin{align}
V_q(Q) & = \Theta(Q - q_\Delta) (E_0 - \Delta - \bar{E}(Q)),
\end{align}
where $\Theta(x)$ is the Heaviside step function and $\bar{E}(Q)$ is a continuous function that linearly interpolates over the energies $E(q)$ of the 1D PES. Due to this interpolation, as well as the nature of the projection process, the corrected potential is not exactly flattened. \\

\begin{figure}[h!]
\includegraphics[width=8.6cm]{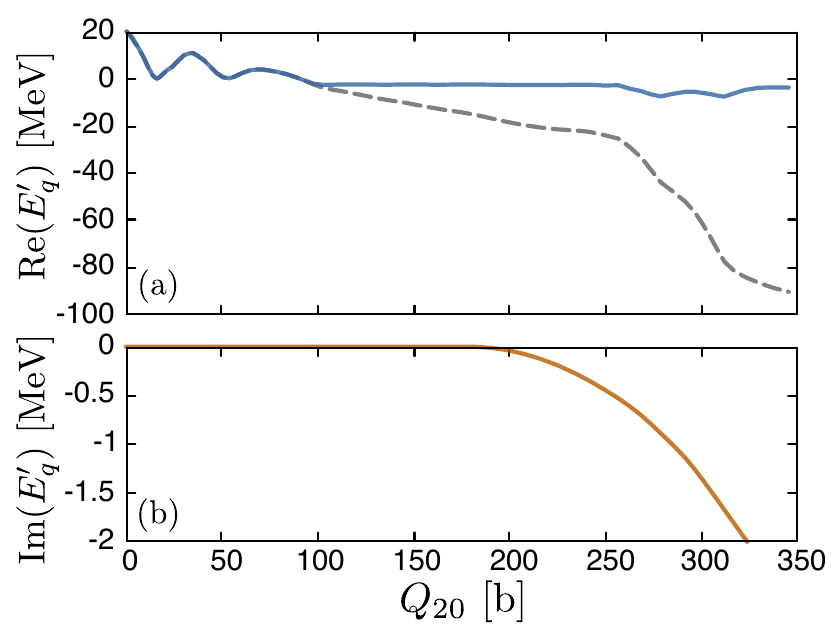}
\caption{\label{fig:256Fm_vprime_abs}The real (solid blue line, upper panel) and imaginary (solid orange line, lower panel) components of the PES after modifications to produce the quasistatic potential. The set of generator states with $\bar{N} = 0.62$ was used for this calculation, with the parameters $r_\mathrm{abs} = -1\xten{-4} \unitsn{MeV \cdot b^{-2}}$, $Q_\mathrm{abs} = 180$~b, and $\Delta = 2$~MeV. The unmodified energy is drawn with a dashed grey line under the real component of the new potential for reference.}
\end{figure}

The absorption potential is defined as a quadratic function that begins at a fixed threshold in the collective coordinate $Q$:
\begin{align}
W(Q) & = r_\mathrm{abs} \Theta(Q - Q_\mathrm{abs}) (Q - Q_\mathrm{abs})^2, \label{eq:absorption_potential}
\end{align}
where the absorption rate $r_\mathrm{abs}$ and threshold $Q_\mathrm{abs}$ are free parameters. The choice of the absorption threshold is physically motivated to be a point past the fission barrier(s) of the 1D PES, ensuring that absorption occurs for nuclear configurations that are approaching or have undergone scission. An example of the real and imaginary components of the modified PES $E'_q(q) = H'(q,q)$ is shown in Figure \ref{fig:256Fm_vprime_abs} to illustrate the shapes of the new potentials. \\

The effects of varying these newly introduced parameters, particularly $r_\mathrm{abs}$, on the final spontaneous fission half-lives are investigated in Section~\ref{sec:results:params_abs}. \\

\begin{figure}[h!]
\includegraphics[width=8.6cm]{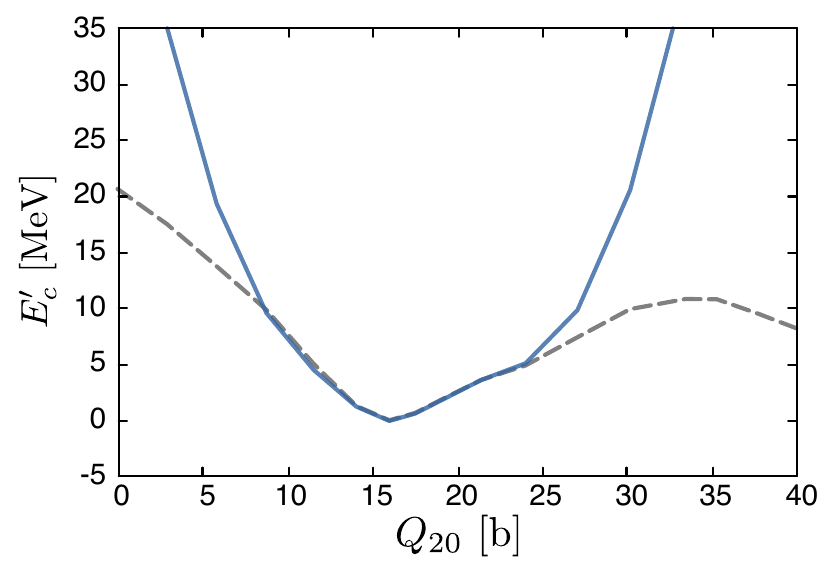}
\caption{\label{fig:256Fm_vprime_conf} The PES after modification to produce the confining potential around the ground state well (solid blue line). This calculation uses the set of generator states with $\bar{N} = 0.62$ and a quadratic coefficient $c = 4\xten{-1} \unitsn{MeV \cdot b^{-2}}$. The unmodified PES is drawn with a dashed grey line for reference.}
\end{figure}

In order to identify the eigenstate of the quasistatic potential which is in resonance with the ground state, a confined potential must be separately created for the set of generator states. The correcting function $V_c(Q)$ for the confining potential is defined to be zero inside the inner part of the ground state well of the PES, while extrapolating with a quadratic outside of it to produce an infinite potential well. The zero region is bounded by the two ``midpoints'' at positions $Q_\mathrm{mid,L}$ and $Q_\mathrm{mid,R}$ which correspond to the energy $E_\mathrm{mid} = (E_0 + V_B)/2$, and the piecewise definition of $V_c(Q)$ is given by
\begin{align}
V_c(Q) & = \begin{cases} 0 & \bar{E}(Q) \leq E_\mathrm{mid} \\ E_\mathrm{mid} - \bar{E}(Q) + c (d_\mathrm{mid})^2 & \bar{E}(Q) > E_\mathrm{mid} \end{cases},
\end{align}
where $d_\mathrm{mid}$ is the distance between $Q$ and the closest midpoint, and $\bar{E}(Q)$ is the interpolated energy of the 1D PES as before. The parameter $c$ of the extrapolating quadratic function is important when seeking the spectrum of highly excited states of the ground state well, but in the current work it is of minimal significance because only the lowest-energy bound states obtained from this potential are of interest. The correcting kernel that is added to the Hamiltonian kernel to form the confining potential is then determined from Eq. \eqref{eq:correcting_kernel} as for the quasistatic case, inserting the modified potential function $V'(Q) = V_c(Q) - E_\mathrm{rot}(Q)$. An example of the resulting confining potential is shown in Figure \ref{fig:256Fm_vprime_conf}.

\subsection{Calculation of bound and quasistatic eigenstates}

Given the modified Hamiltonian kernels $H'(q,q')$ defined in the previous subsection, the TDGCM formalism outlined in Section \ref{sec:methodology:tdgcm} can be straightforwardly followed up to Eq. \eqref{eq:coll_hamiltonian} to obtain the collective Hamiltonian $H_C(a,a')$ for each modified potential. To obtain the eigenstates for each potential, one considers the time-independent analogue of Eq. \eqref{eq:coll_schrodinger2}, which is simply
\begin{align}
H_C \cdot \bv{g} & = E \bv{g}.
\end{align}
The diagonalisation and eigendecomposition of $H_C$ straightforwardly produces
\begin{align}
H_C \cdot \bv{v}_i & = E_i \bv{v}_i,
\end{align}
where each eigenvector $\bv{v}_i$ is an energy eigenstate of the collective Hamiltonian, with energy $E_i$, that can be represented in the natural basis as
\begin{align}
\ket{\psi_i} & = \int da\ v_i(a) \ket{\psi(a)}\!.
\end{align}
In order to help with the identification of the lowest-energy eigenstates of the ground state well needed for spontaneous fission calculations, the expectation values and standard deviations of $Q_{20}$ for each eigenstate can be calculated using the projection kernels according to
\begin{align}
\langle \hat{Q}_{20} \rangle \equiv \bra{\psi_i} {\hat{Q}_{20}} \ket{\psi_i} & = \frac{\sum_Q \bra{\psi_i} \hat{P}^L_{\hat{Q}}(Q) \ket{\psi_i} Q}{\sum_Q \bra{\psi_i} \hat{P}^L_{\hat{Q}}(Q) \ket{\psi_i}},
\end{align}
where each sum includes all eigenvalues of $\hat{Q}$ allowed in the discretised projection mesh defined alongside the projector in Section~\ref{sec:methodology:discrete_projection}; $\langle \hat{Q}_{20}^2 \rangle$ can be obtained analogously by squaring the factor of $Q$ in the numerator. \\

\begin{figure}[h!]
\includegraphics[width=8.6cm]{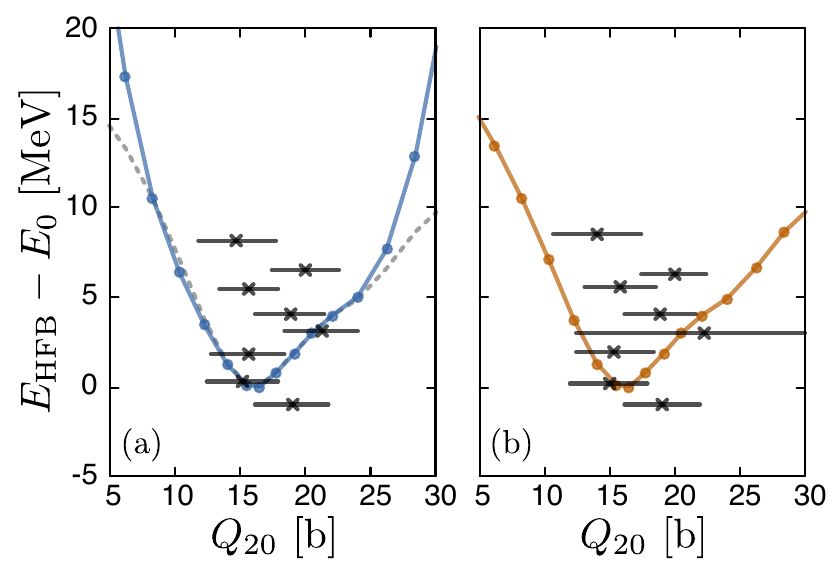}
\caption{\label{fig:eigenstates_256Fm_conf_qs} Positions in $Q_{20}$ and energies of the eigenstates of the confined potential (a) and quasistatic potential (b), calculated using the $\elem{256}{}{Fm}$ dataset with average overlap $\bar{N} = 0.78$. In (a), the original 1D PES is drawn with a black dotted line, while the confining potential is marked by a solid blue line with circle markers; in (b), the quasistatic potential (which matches the original PES in this region) is plotted using a solid orange line with circle markers. The black crosses mark the energies and $\langle\hat{Q}_{20}\rangle$ values of the eigenstates, with horizontal error bars indicating their standard deviation in $\hat{Q}_{20}$ as a $1\sigma$ interval. Eigenstates with energies higher than $E_0 + 10$~MeV or with $\langle\hat{Q}_{20}\rangle > 30$~b are excluded from both plots for clarity.}
\end{figure}

Once the eigenstates' positions in $Q_{20}$ are well quantified, they can be plotted against their (real) energies $E_i$ for both the confined and quasistatic potentials. The eigenstates for the data set with average overlap $\bar{N} = 0.78$ is given in Figure \ref{fig:eigenstates_256Fm_conf_qs}, displaying the states of the two modified potentials side-by-side for comparison. It is immediately evident that the ground state and the first two excited states of each potential agree very closely in their energies and $Q_{20}$ positions, while subsequent excited states differ by increasing amounts due to the changing shape of the two potentials at higher energies. The practically exact match in the real energies of the lowest three quasistatic eigenstates in Fig.~\ref{fig:eigenstates_256Fm_conf_qs}(b) to the corresponding bound states in Fig.~\ref{fig:eigenstates_256Fm_conf_qs}(a) is sufficient to conclude that the former are quasistatic versions of the latter coupled to the ``continuum'' that simulates the decay channel of spontaneous fission. As an additional measure of certainty, the overlaps $\braket{\psi^q_i}{\psi^c_i}$ between the paired eigenstates are also computed to verify that they approach 1. \\

Once the quasistatic state in resonance with the ground state of the bound potential is identified, it is a straightforward matter to calculate the spontaneous fission half-life $t_{1/2}$ from the imaginary component of the state's energy as explained in Section~\ref{sec:methodology:quasistatic}.


\section{\label{sec:results}Results and discussion}

\subsection{Spontaneous fission half-lives}

The new modelling methods described in Section \ref{sec:applications} have been applied primarily to the one-dimensional PES of $\elem{256}{}{Fm}$ in order to calculate its spontaneous fission half-life and compare it with experimental values. However, the components of the model introduce a range of parameters which cannot be defined from physical arguments. In order for the predicted half-lives to hold any measure of validity, the variation of results with respect to these parameters must first be carefully examined.

\subsubsection{\label{sec:results:params_abs}Parameters of the absorption potential}

The absorption rate $r_\mathrm{abs}$ is directly related to the spontaneous fission half-lives obtained with the method, because the imaginary potential it controls is responsible for the imaginary component of the energy eigenvalue of the quasistatic state to which the spontaneous fission half-life is inversely proportional. However, this proportionality should not characterise the general relationship between the half-life and the absorption rate. The qualitative impact of an imaginary potential on the dynamics of a quantum system is the presence of a ``sink'' which absorbs (removes) the wavefunction within its region of influence\footnote{A positive imaginary potential produces a ``source'' with the opposite effect, but the absorption rate is strictly negative in this work.}. An appropriate imaginary potential should then simulate an open boundary condition and prevent reflections from the finite boundaries of the calculation domain. If the other choices made to define the absorption potential ($Q_\mathrm{abs}$ and the functional form) are appropriate, calculating and plotting the spontaneous fission half-life as a function of $r_\mathrm{abs}$ will reveal an interval of values for $r_\mathrm{abs}$ where the overall absorption is sufficiently well-tuned to prevent reflections. \\

\begin{figure}[t]
\includegraphics[width=8.6cm]{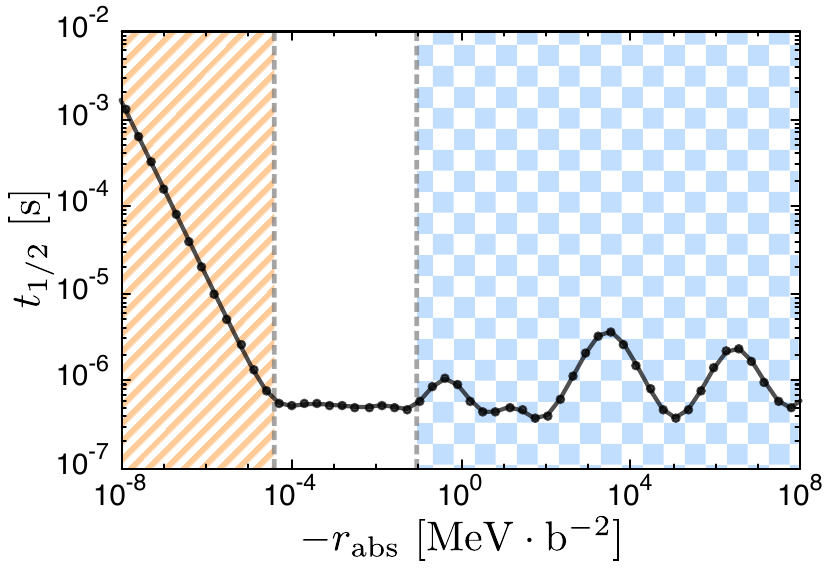}
\caption{\label{fig:sf_lifetime_256Fm_7s} Spontaneous fission half-lives $t_{1/2}$ of $\elem{256}{}{Fm}$ as a function of the negative absorption rate $-r_\mathrm{abs}$, using a basis of generator states with overlaps $\bar{N} = 0.62$. Three distinct regimes of behaviour with respect to $-r_\mathrm{abs}$ are highlighted: regions shaded with orange stripe (blue checkerboard) patterns indicate when the absorption strength is too low (high); between these two regimes, the unshaded region is the interval in $-r_\mathrm{abs}$ where the spontaneous fission half-life remains relatively constant.}
\end{figure}

The relation between $r_\mathrm{abs}$ and $t_{1/2}$ is shown in Figure \ref{fig:sf_lifetime_256Fm_7s}. When $r_\mathrm{abs}$ is small, the spontaneous fission half-life $t_{1/2}$ follows an inversely linear relation, while at high absorption rates it oscillates somewhat randomly. However, between these two regimes of behaviour there is an interval spanning several orders of magnitude in $r_\mathrm{abs}$ where the spontaneous fission half-life remains roughly constant. This range of ``convergence'' can be interpreted as the region where the absorption strength is well chosen to simulate an open boundary condition. These different regimes of behaviour reliably appear when surveys of $t_{1/2}$ as a function of $r_\mathrm{abs}$ are performed for the quasistatic ``ground'' state obtained for all different mesh densities and nuclei studied in this work. \\

\begin{figure}[h!]
\includegraphics[width=8.6cm]{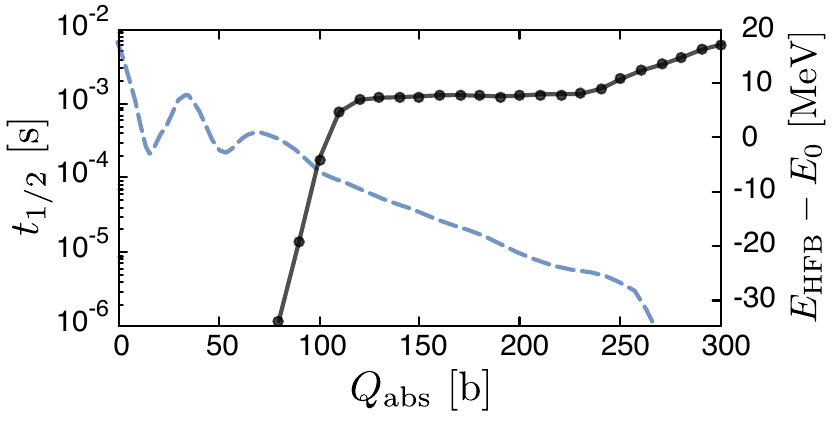}
\caption{\label{fig:sf_lifetime_q_256Fm_7s} Spontaneous fission half-lives $t_{1/2}$ of $\elem{256}{}{Fm}$ as a function of the absorption threshold $Q_\mathrm{abs}$, using the data set with $\bar{N} = 0.62$ as for Figure \ref{fig:sf_lifetime_256Fm_7s}. The black line with dot markers indicates the change in $t_{1/2}$ as $Q_\mathrm{abs}$ is shifted in the $Q_{20}$ dimension, for a fixed absorption rate $r_\mathrm{abs} = -10^{-3} \unitsn{MeV \cdot b^{-2}}$. The original PES is plotted with a dashed blue line (with energies indicated by the right axis) for reference.}
\end{figure}

The absorption potential given in Eq. \eqref{eq:absorption_potential} is also parametrised by the value $Q_\mathrm{abs}$, which determines the point in $Q$ beyond which the imaginary potential is applied. Provided that the absorption of the wavefunction occurs in the region of the PES after the fission barrier(s) where the real potential is flattened and that the interval between $Q_\mathrm{abs}$ and the maximum value of $Q$ in the domain is large enough for the absorption to take effect, the dynamics are relatively insensitive to its variation. This can be straightforwardly proven by plotting $t_{1/2}$ as a function of $Q_\mathrm{abs}$, as shown in Figure \ref{fig:sf_lifetime_q_256Fm_7s}. It is obvious that the variation in the spontaneous fission half-life is minimal for a large range of $Q_\mathrm{abs}$ values between 130 and 230~b. Deviations occur when the absorption threshold is either placed too close to the second fission barrier ($Q_{20} < 130$~b), or too close to the boundary of the projection mesh (at $Q_{20} > 230$~b) such that the absorption is insufficient to prevent reflections. It is safe to conclude that there is a generous range within which $Q_\mathrm{abs}$ may be chosen without any significant impact on the resulting value of $t_{1/2}$. \\

The above results demonstrate that the variation of both $r_\mathrm{abs}$ and $Q_\mathrm{abs}$ can affect the calculated spontaneous fission half-lives, but there are intervals for each parameter where $t_{1/2}$ ``converges'' with respect to each. Because the convergence range for $Q_\mathrm{abs}$ is large and depends on the shape of the PES, it is assumed that a single value of $Q_\mathrm{abs}$ can be chosen for all calculations. However, the convergence in $r_\mathrm{abs}$ is less certain; it is desirable to compute a statistical average for $t_{1/2}$ as well as a measure of the quality of its convergence. To do this in a programmatic fashion, the ``convergence interval'' is defined for any absorption rate $r \equiv r_\mathrm{abs}$ as the largest interval $[r_\mathrm{min}, r_\mathrm{max}]$ such that
\begin{align}
\frac{1}{\eta} \leq \frac{t_{1/2}(r')}{t_{1/2}(r)} \leq \eta
\end{align}
for all $r' \in [r_\mathrm{min}, r_\mathrm{max}]$ and a constant $\eta > 1$ which is set to 1.1 in the current work. The convergence range for $r$ is defined as $R_c(r) = r_\mathrm{max} / r_\mathrm{min}$. The mean of the spontaneous fission half-life $\bar{t}_{1/2}(r)$ is then calculated within the region of convergence according to
\begin{align}
\bar{t}_{1/2} & = \frac{1}{\mathcal{N}} \sum_{r'=r_\mathrm{min}}^{r_\mathrm{max}} t_{1/2}(r'),
\end{align}
where $\mathcal{N}$ is the number of discrete elements in the interval $[r_\mathrm{min},r_\mathrm{max}]$. The final spontaneous fission half-life is then the value of $\bar{t}_{1/2}(r)$ corresponding to the absorption rate $r$ with the largest convergence range; in the case of a tie for the largest $R_c$, the smallest $\bar{t}_{1/2}(r)$ is accepted. This prescription eliminates the $r_\mathrm{abs}$ ``degree of freedom'' from the outcome and produces a single value of the spontaneous fission half-life, but also provides the measure of convergence $R_c$ which serves as a warning flag in the case of poorly-converged results.

\subsubsection{\label{sec:results:mesh_density}Density of the set of generator states}

As described in Section \ref{sec:applications:1d_fission_path}, the procedure of choosing a sparse set of generator states (as opposed to truncating eigenstates with small eigenvalues from the natural basis) has not been widely attempted in the literature \cite{martinezlarraz2022,lauphdthesis}. In order to investigate the consequences of this new approach, calculations were repeated for the various data sets defined in Table \ref{tab:mesh_densities} in order to search for spurious or unexpected behaviours that might indicate a mesh density that was too low or too high. The spontaneous fission half-lives obtained for these varying mesh densities were determined using a fixed $Q_\mathrm{abs} = 180$~b while statistically averaging over the converged values of $r_\mathrm{abs}$ as described above. They are compared below in Figure \ref{fig:sf_lifetime_256Fm_cmp}(a), alongside the maximum convergence range $R_c$ for each data set in Fig. \ref{fig:sf_lifetime_256Fm_cmp}(b). Since the sampling of $r_\mathrm{abs}$ is performed exponentially in powers of 2, the values of the convergence ranges are discrete. \\

\begin{figure}[h!]
\includegraphics[width=8.6cm]{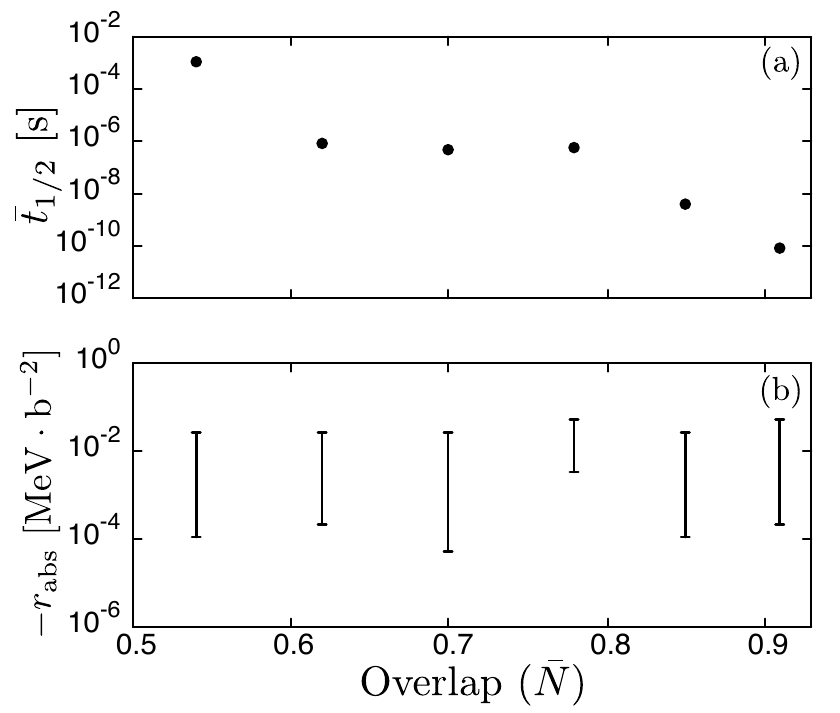}
\caption{\label{fig:sf_lifetime_256Fm_cmp}A comparison of the spontaneous fission half-lives obtained from sets of input wavefunctions with different mesh densities (represented on the $x$-axis by the average overlap between neighbouring states in the set): (a) shows the mean half-life $\bar{t}_{1/2}$, averaged over the maximal convergence range, while (b) depicts the extent of each convergence range in $r_\mathrm{abs}$. See main text for details.}
\end{figure}

The general trend visible in Fig. \ref{fig:sf_lifetime_256Fm_cmp}(a) is a decrease in $\tau_\mathrm{SF}$ as the density of the set of generator states increases. However, given that the $y$-axis in this plot is logarithmic, these variations with overlap span many orders of magnitude. The sizes of the convergence regions in Fig. \ref{fig:sf_lifetime_256Fm_cmp}(b) generally do not suggest any problems with the stability of $t_{1/2}$ with respect to $r_\mathrm{abs}$ (except in the case of the $\bar{N} \sim 0.78$ data set, whose convergence range is smaller). In order to explain these differences in results and decide whether there is a particular overlap $\bar{N}$ (or range of overlaps) that produces the most ``correct'' spontaneous fission half-lives, it is necessary to consider in more detail the possible impacts of changing the mesh density on the model. \\

\begin{figure}[h!]
\includegraphics[width=8.6cm]{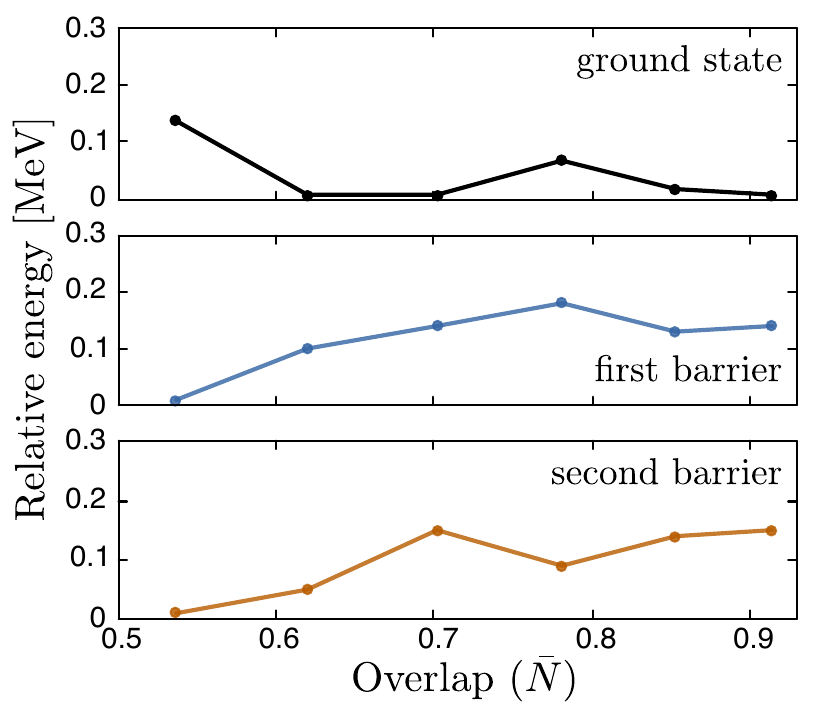}
\caption{\label{fig:gs_barriers_256Fm_cmp} Differences in the energies of the HFB wavefunctions identified as the ground state (top), and as the peaks of the first barrier (middle) and second barrier (bottom) when the mesh density is varied. The energies for each plot have been arbitrarily shifted to facilitate comparisons between data sets with different overlaps.}
\end{figure}

It is self-evident that if the mesh density of the PES is not sufficiently high, it will not give an accurate representation of the energetic landscape of the fissioning nucleus. The resulting lack of detail is likely to affect the heights and shapes of the fission barriers, to which spontaneous fission half-lives are critically sensitive. The variations in the energies of the ground state and barriers between the different data sets are depicted in Figure \ref{fig:gs_barriers_256Fm_cmp}. In general, when a sparser mesh is used to select the data set (corresponding to lower overlaps between neighbours), the energy of the ground state increases while that of the barrier states decreases; this is because the larger intervals between states make it less likely that the path reaches the true energy minimum (for the ground state) or maximum (for the barrier states). However the largest variation in barrier heights across the data set is less than 0.2 MeV, which is insignificant compared to differences induced by changing the nucleon-nucleon interaction or the size of the harmonic oscillator basis. Furthermore, the qualitative effect of a reduced barrier height with a sparser mesh should be the shortening of spontaneous fission half-lives, while the opposite trend is observed in Figure \ref{fig:sf_lifetime_256Fm_cmp}(a). It can therefore be surmised that the impact on the final result of the variation of barriers with respect to mesh density is minimal. \\

\begin{figure}[h!]
\includegraphics[width=8.6cm]{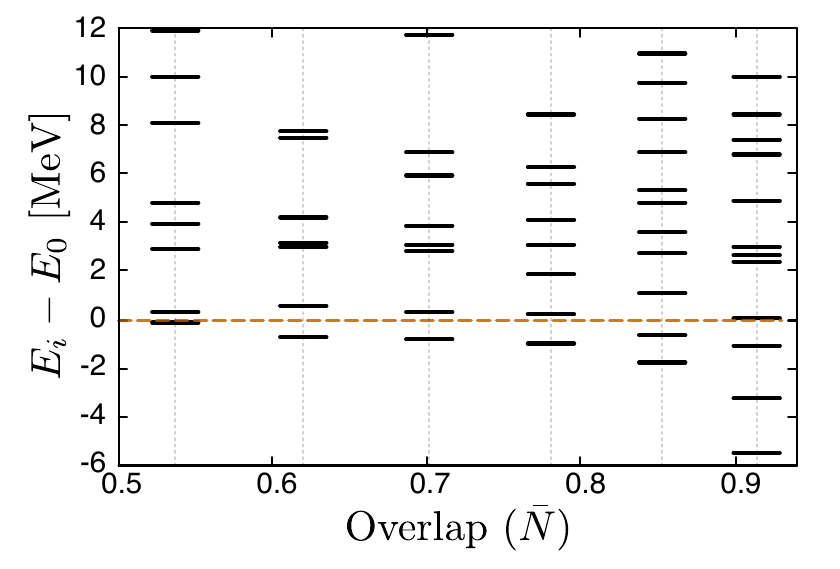}
\caption{\label{fig:qs_energies_256Fm_cmp}The energies of the lowest quasistatic states, which are the eigenstates of the kernel representing the Hamiltonian given in Eq. \eqref{eq:quasistatic_hamiltonian}. Different data sets, differentiated by $\bar{N}$, produce different spectra of eigenstates. The reference energy $E_0$ (orange dashed line) is the energy of the HFB wavefunction identified as the ground state in the full 1D PES. Eigenstates whose average $Q_{20}$ values lie outside of the ground state well are excluded; see main text for details.}
\end{figure}

A different and more complex issue is revealed by considering the energies of the eigenstates of the bound system that is used to identify resonances in the quasistatic approach. Because of the discrete nature of the calculation, the number of eigenstates obtained is equal to the number of generator states; however, one should expect the overall shape of the spectrum not to change significantly with the mesh density. The energies of the eigenstates for the various mesh densities studied are compared in Figure \ref{fig:qs_energies_256Fm_cmp}. To avoid cluttering the spectrum with irrelevant states, the plot only includes the eigenstates in the ``ground state well''; these states have an expectation value of $Q_{20}$ less than 35~b (the approximate elongation of the first barrier). While there is a significant degree of variation in the energy spectrum as a function of the overlap, the low-lying structure of the spectrum remains fairly similar across the data sets with average overlaps of 0.62, 0.70 and 0.78. As the mesh density decreases, it is also to be expected that the worsening representation of the ground state well and the shrinking basis of HFB states alters the spectrum as found for the sparsest set of states with $\bar{N} = 0.53$. However, the drastic changes in the spectrum and lowering of overall energies for overlaps greater than 0.8 are signs of unstable behaviour which must be considered in greater detail. \\

\begin{figure}[h!]
\includegraphics[width=8.6cm]{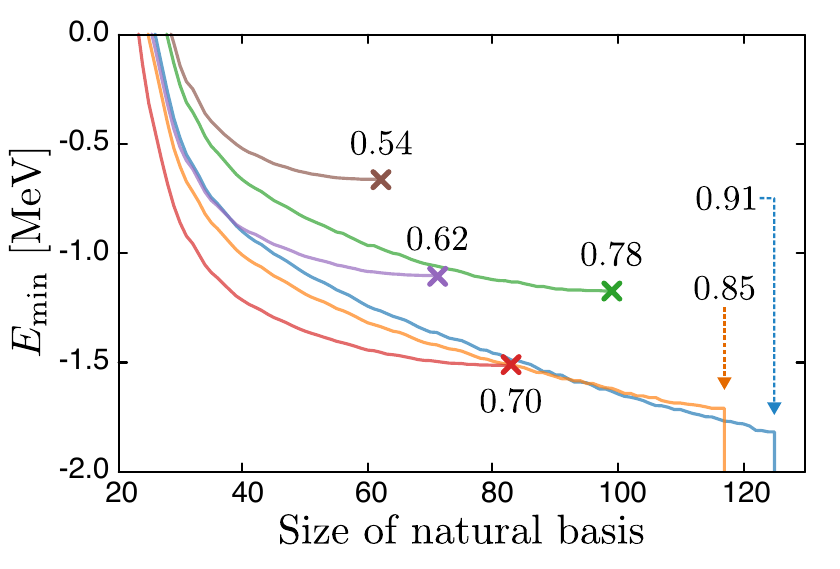}
\caption{\label{fig:256Fm_nbasis_energies}The energy $E_\mathrm{min}$ of the lowest eigenstate (relative to the ground state energy of the PES) as a function of the natural basis size for the various data sets. The largest basis size for each data set, where no states are truncated from the natural basis, is marked with a cross and labelled by its average overlap $\bar{N}$ (this point occurs below the minimum displayed energy for the $\bar{N} = 0.85$ and $0.91$ cases).}
\end{figure}

The ``collapse'' of the eigenstate energies is reminiscent of behaviour observed when the number of states in the natural basis of GCM calculations is too large. Articles by Bonche \textit{et al.} \cite{bonche1990} and more recently by Mart\'{\i}nez-Larraz and Rodr\'{\i}guez \cite{martinezlarraz2022} both plot the energies of the eigenstates of GCM calculations as functions of the number of states retained in (or equivalently, truncated from) the natural basis; the latter also considers multiple nuclei and sets of generator states with different densities. In both works, when the size of the natural basis exceeds a particular threshold the energy of the lowest eigenstate abruptly plummets by several MeV (the exact decreases are not shown). By following the natural basis formalism given in Eq. (8) of Ref. \cite{martinezlarraz2022}\footnote{This is the same definition set out in Eq. (10.14) of Ref. \cite{ringandschuck}.}, equivalent calculations were performed to determine the lowest eigenstate energy as a function of the size of the natural basis for all of the data sets in Table~\ref{tab:mesh_densities}. The results of this process are shown in Figure \ref{fig:256Fm_nbasis_energies}. \\

Comparing these results with Fig. 1(i) of Ref. \cite{martinezlarraz2022} leads to the deduction that the high mesh density of the $\bar{N} = 0.85$ and $\bar{N} = 0.91$ data sets induces the instability of the natural basis responsible for the collapse in the lowest eigenstate energy; this is solid evidence that the results obtained in these two cases are affected by numerical issues and should be discarded, or at least treated with scepticism. \\

Having said this, it should not be overlooked that there is another potential explanation for the phenomenon of ``eigenspectrum collapse'' in the GCM when using a dense set of generator states. Bender, Duguet, Lacroix \textit{et al.} \cite{lacroix2009,bender2009,duguet2009} published a series of articles highlighting the issue of self-interaction and self-pairing contributions to the Hamiltonian kernel. These erroneous terms arise from two sources: the unequal treatment of direct and exchange components of the nucleon-nucleon interaction, and the handling of density-dependent components when replacing the Hamiltonian with an energy density functional according to density functional theory. These factors produce spurious contributions towards the mean-field solution of the HFB equations used to construct a static PES, but also lead to divergences in the off-diagonal terms of the Hamiltonian kernel in the GCM and TDGCM. One way to solve most, if not all, of the problems arising from self-interaction and self-pairing is to include the exact pairing and exchange components of the Gogny force in HFB calculations, as demonstrated in \cite{anguiano2001}; such approach may be considered in future investigations depending on the additional computational complexity it entails. \\

Given this subtle and fundamental concern, further research is needed to convincingly decide whether the numerical precision issues or the self-interaction and self-pairing problems are to blame for the spurious behaviours revealed by Figures \ref{fig:qs_energies_256Fm_cmp} and \ref{fig:256Fm_nbasis_energies}. Nevertheless, the following conclusions can still be drawn with regard to the mesh density:
\begin{itemize}
\item A mesh of generator states that is too sparse will degrade the fidelity of the PES's representation in the basis of the calculations.
\item A mesh of generator states that is too dense produces numerical instabilities when constructing the natural basis.
\end{itemize}
In excluding the results produced by the two overly-dense data sets, Figure \ref{fig:sf_lifetime_256Fm_cmp}(a) shows that the spontaneous half-life decreases asymptotically as the average overlap increases. The spontaneous fission half-life is taken to be the mean of the results of the $\bar{N} \sim 0.62$, 0.70, and 0.78 data sets, which appear ``converged'' while being free of instabilities according to Fig.~\ref{fig:256Fm_nbasis_energies}, giving a final result of $t_{1/2} = 8.318 \xten{-7}$ s.


\subsection{\label{sec:results:proj_prob_visualisation}Projected probabilities for visualisation}

A final method to verify the sanity of the results obtained above is to visualise the wavefunctions involved in the calculations. Although such analysis is highly qualitative, it is nevertheless useful in this context because the degree of freedom $Q_{20}$ that parametrises the collective nuclear shape can be interpreted as a spacelike coordinate; the dynamics of the compound nucleus moving across the 1D PES should then resemble those of a quantum mechanical particle in a potential. This provides strong intuitive predictions for the shape of the wavefunction with respect to $Q_{20}$. One measure of the probability to observe the nuclear wavefunction $\ket{\Psi(t)}$ to be occupying the generator state $\ket{\phi(q)}$ \cite{verriere2020,deng2026} is simply the norm squared of the product of the two:
\begin{align}
\mathrm{Pr}\big(\!\ket{\phi(q)}\!,t\big) & = \big|\!\braket{\phi(q)}{\Psi(t)}\!\big|^2 \notag \\
& = \bigg|\int_\mathcal{Q} dq'\ f(q',t) N(q,q')\bigg|^2. \label{eq:probability_overlap}
\end{align}
However, this ``overlap-based probability'' does not represent the likelihood for the nucleus to actually have a collective shape corresponding to the generator coordinate $q$ because the latter is only the expectation value of the generator state $\ket{\phi(q)}$. It can also be easily proven that the total probability according to this definition is not normalised. \\

 Projection onto the operators of the generator coordinates as given in Eq. \eqref{eq:projected_probability} provides a natural and rigorous solution to this conundrum. Having calculated the probability kernels beforehand, the probability of observing $\ket{\Psi(t)}$ to have the collective coordinate $Q$ (associated with the observable operator $\hat{Q}$) at time $t$ is given by
\begin{align}
\mathrm{Pr}(Q,t) & = \bra{\Psi(t)} \hat{P}^L_{\hat{Q}}(Q) \ket{\Psi(t)} \notag \\
& = \iint_{\mathcal{Q}^2}\!\! dq\,dq'\, f^*(q,t) f(q',t) P^L_{\hat{Q}}(Q;\,q,q'). \label{eq:projected_probability_time}
\end{align}
This ``projected probability'' is exact and independent of the choice of generator states, and it is defined for each of the $L$ mesh points sampled in the eigenvalue spectrum of the projected operator $\hat{Q}$. By contrast, the overlap-based definition given in Eq. \eqref{eq:probability_overlap} only yields a probability for each value of $q$ represented in the set of generator states. \\

It should be noted that, because of the discretisation of the collective coordinate $Q_{20}$ and its corresponding operator matrix in Eq.~\eqref{eq:q20_kernel}, the probability distribution associated with this quantity is not properly normalised to 1. In future work, this shortcoming will be alleviated by increasing the number of shells in the HO basis or by transforming the entire problem into three-dimensional Cartesian space. \\

\begin{figure}[h!]
\includegraphics[width=8.6cm]{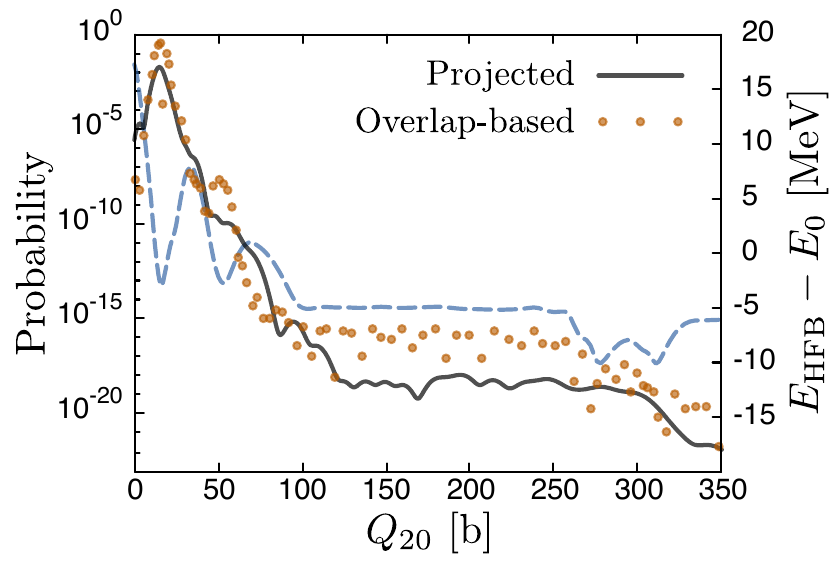}
\caption{\label{fig:256Fm_q20_pprob_weight_cmp} Comparison between the wavefunction's projected probability $\mathrm{Pr}(Q)$ (Eq. \eqref{eq:projected_probability_time}, solid black line) and the overlap-based probability $\mathrm{Pr}(|\phi(q)\rangle)$ (Eq. \eqref{eq:probability_overlap}, orange circles). Both quantities are plotted with respect to the left $y$-axis as functions of $Q_{20}$. The real part of the quasistatic potential is plotted for reference according to the right $y$-axis with a dashed blue line. See main text for details.}
\end{figure}

While the projected probability can be calculated for any GCM or TDGCM wavefunction, in the present work it is used to examine the quasistatic states from which the spontaneous fission half-life is derived. The relevant quasistatic state for the $\bar{N} = 0.70$ data set is plotted in Figure \ref{fig:256Fm_q20_pprob_weight_cmp}, comparing the overlap-based probability $\mathrm{Pr}(|\phi(q)\rangle)$ (orange circles) with the projected probability $\mathrm{Pr}(Q)$ (solid black line), both of which are normalised \textit{a posteriori} to ensure fairness. \\

Both definitions are capable of demonstrating the expected behaviour with respect to the quasistatic potential (dashed blue line): the nuclear system remains largely confined to the ground state well, but quantum tunnelling permits it to propagate through the barriers with an exponential decay in occupation probability until reaching the flattened part of the potential. The effect of the potential well of the isomer state at $Q_{20} \sim 52$~b is also noticeable as a bump in both probabilities. Overall, the qualitative properties of the two types of probability are similar. However, given that the overlap-based probability is not normalised by definition and has a resolution dependent on the size of the set of generator states, the projection-based probability should be preferred as a more rigorous and meaningful alternative.

\subsection{Comparison with experimental values}

In addition to the detailed analysis presented in this Section for the spontaneous fission of $\elem{256}{}{Fm}$, calculations were also performed on the 1D PES of $\elem{256}{}{Cf}$, using the methods explained in Section \ref{sec:applications:1d_fission_path} with an overlap constraint of 0.70. The resulting spontaneous fission half-lives are compared to experimental values according to the IAEA's online chart of nuclides \cite{iaeachart} in Table \ref{tab:sf_lifetimes}. \\

\begin{table}[h!]
\begin{tabular}{c|c|c|c}
Nuclide & $t_{1/2}^\mathrm{th}$ [s] & $t_{1/2}^\mathrm{exp}$ [s] & $t_{1/2}^\mathrm{th} / t_{1/2}^\mathrm{exp}$ \\
\hline
$\elem{256}{}{Fm}$ & $8.318 \xten{-7}$ & $9.426 \xten{3}$ & $8.825 \xten{-11}$ \\
$\elem{256}{}{Cf}$ & $2.287 \xten{-12}$ & $7.38 \xten{2}$ & $3.10 \xten{-15}$
\end{tabular}
\caption{\label{tab:sf_lifetimes}A comparison between the spontaneous fission half-lives $t_{1/2}^\mathrm{th}$ obtained in this work and their experimentally measured counterparts $t_{1/2}^\mathrm{exp}$ \cite{iaeachart} for the two nuclides $\elem{256}{}{Fm}$ and $\elem{256}{}{Cf}$. See main text for discussion.}
\end{table}

The theoretical spontaneous fission half-lives calculated in the present work are much shorter than the experimentally observed values. However, it should not be forgotten that no spontaneous fission models are currently capable of achieving quantitative agreements with experimental observations. Table \ref{tab:256Fm_lifetimes_cmp} presents a range of predictions for the spontaneous fission half-life of $\elem{256}{}{Fm}$ obtained by various WKB-based simulation methods. Different choices of nuclear interaction, approximation for the collective inertia, and phenomenological parameters such as pairing strength lead to variations of results spanning many orders of magnitude. It is interesting to note that the majority of these other models tend to produce half-lives which are longer than the experimental value, whereas the microscopic approach of this work displays the opposite behaviour. \\

\begin{table}[h!]
\begin{tabular}{c|c}
Source & $t_{1/2}$ [s] \\
\hline
Experiment \cite{iaeachart} & $9.426 \xten{3}$ \\
This work & $8.318 \xten{-7}$ \\
Baran 2015 \cite{baran2015} & $\sim 10^5$ \\
Giuliani 2013 \cite{giuliani2013} & $\sim 10^{-7}$ to $\sim 1$ \\
Warda 2012 \cite{warda2012} & $1.2 \xten{7}$ \\
Erler 2012 \cite{erler2012} & $\sim 10$ to $\sim 10^{12}$
\end{tabular}
\caption{\label{tab:256Fm_lifetimes_cmp}A comparison of the experimental spontaneous fission half-lives of $\elem{256}{}{Fm}$ with the calculated values from a range of published theoretical models. The $\sim$ symbol indicates that the cited values are only represented graphically in the original source and are hence approximate.}
\end{table}

In any case, further refinement of this model, concerning both theory and implementation, is needed to improve its observable outputs. In this regard, two significant factors which are commonly discussed within the sphere of the semiclassical WKB method are relevant: the collective inertia and the choice of effective interaction. \\

The collective inertia is a purely phenomenological property that describes the effective mass of the nuclear system as a function of the collective coordinates involved; it is a key component of the action which is minimised to obtain the semiclassical spontaneous fission path.  However, it is well documented \cite{schunck2016,ringandschuck,peierls1962} that the values of the collective inertia produced by a GCM-based approach are erroneously small; the inertias given by an approach using adiabatic time-dependent Hartree-Fock-Bogoliubov (ATDHFB) theory are strongly preferred. Analysis of this problem has revealed that, in order to produce correct collective inertias, the generator coordinates of the GCM formalism must include not only space-like generator coordinates $\bv{q}$ but also their associated momenta $\bv{p}$ \cite{peierls1962}, resulting in the generalised ``dynamic'' GCM (DGCM) proposed in \cite{goeke1980}. Even though the collective inertia is never explicitly defined in the current microscopic approach, there is a tangible possibility that the shortcomings of the GCM are affecting the results in a similar way. Implementation of the DGCM to obtain the correct collective inertias has been performed for light nuclei \cite{hizawa2022}, but considerable optimisation and/or additional computing capacity are still needed before the technique is suitable for reactions on the scale of nuclear fission. Further complicating this problem is the fact that collective inertias have historically been calculated using the (Inglis) cranking approximation (with an optional perturbative expansion for the momentum operator) \cite{baran2015,giuliani2018} in order to avoid the expensive matrix inversions involved in the full numerical solution. Various studies have been conducted to compare the collective inertias obtained with different approximations, presenting differing conclusions on the suitability of the cranking formalism \cite{sadhukhan2014,washiyama2021}. Of particular note, Giuliani and Robledo found in \cite{giuliani2018} that lessening the degree of simplification reduces the discrepancies between the inertias obtained with the GCM and ATDHFB formalisms. \\

The second key factor affecting both the new microscopic approach and existing semiclassical models of spontaneous fission is the effective potential used to describe nucleon-nucleon interactions. The Gogny interaction has a wide variety of accepted parametrisations, fitted to different nuclear data and hence carrying different advantages and disadvantages in their applications. While the D1S parametrisation of the Gogny force used in this work is generally recognised as a good choice across the nuclear chart close to the valley of stability \cite{robledo2019}, various authors have presented arguments in favour of selecting other parametrisations for spontaneous fission: the D1M parametrisation is a successor of D1S which is more suited to neutron-rich, superheavy nuclei \cite{rodriguezguzman2014}, while the tensor components of the interaction introduced by the recent D1ST2a parametrisation significantly alter the fission PESs and resulting spontaneous fission half-lives of actinides \cite{rodriguezguzman2024}. Studies of spontaneous fission have also been carried out with other interactions such as the Barcelona-Catalana-Paris-Madrid energy density functional \cite{giuliani2013}. \\

The prospects of extending the proposed model with DGCM formalism, or of analysing its behaviour with different interactions and parametrisations, lie beyond the scope of the current work. However, the potential impact of correcting the GCM can be seen qualitatively by considering the differences in spontaneous fission half-lives observed in semiclassical models when comparing the use of GCM and ATDHFB inertias, such as the systematic study performed by Rodr\'iguez-Guzm\'an and Robledo \cite{rodriguezguzman2014}, of which Figure 5 presents the calculated spontaneous fission half-lives of a range of actinide and post-actinide nuclides calculated using the WKB formalism and the Gogny D1M interaction. Depending on the initial energy and the compound nucleus, the spontaneous fission half-lives of the actinides are up to $10^{12}$ times longer when using the ATDHFB inertias as opposed to their GCM counterparts. This finding is strongly encouraging for the present work, as increasing the half-lives obtained in Table \ref{tab:sf_lifetimes} by such orders of magnitude would bring them tantalisingly close to experimental values.


\section{\label{sec:conclusion}Conclusion}

This work presents a new method to calculate the spontaneous fission half-life of a nucleus from its one-dimensional PES, combining the existing ``quasistatic approach'' proposed in Ref. \cite{scamps2015} with a modernised version of the TDGCM without the GOA. \\

A key foundation of this method is the novel application of projection techniques to the generator coordinates themselves, which in this case parametrise the collective nuclear shape. The expectation values of these coordinates are frequently constrained to obtain the self-consistent mean-field wavefunctions that form a PES, but the degrees of freedom themselves are not well-defined observables of the nucleus. Obtaining the explicit distributions of the wavefunctions in terms of the generator coordinate $Q_{20}$ using a discretised projection operator leads to the definition of a ``probability kernel'', which allows for the calculation of the probability to observe a wavefunction of the nuclear system (which may be a mixture of mean-field generator states) with a particular collective shape. While this outcome is already useful for effectively visualising and analysing the shapes of wavefunctions, it also provides a rigorous mechanism to apply modifications to the nuclear potential represented by the Hamiltonian kernel. This final application is necessary for the quasistatic approach, which describes spontaneous fission as the decay of a metastable eigenstate obtained from an altered complex potential. \\

The results presented in this article include a thorough examination of the model inputs and other factors that influence the spontaneous fission half-lives produced with the new method. The parameters which control the modification of the potential are shown either to have little impact on the outcome or to display an ``interval of convergence'' over which the result varies minimally. On the other hand, the ``mesh density'' defined by the choice of generator states has more complex effects on the predicted spontaneous fission half-lives. Analysing results across a range of mesh densities leads to the general conclusion that choosing too few or too many generator states will have adverse effects on the subsequent observables of the system. Further investigation beyond the scope of the current work is needed to properly identify and address the underlying causes of this behaviour, but doing so is necessary to ensure that GCM and TDGCM are robustly formulated and correctly applied. \\

Although the final spontaneous fission half-lives obtained by the new model are much shorter than experimental values, a generalisation of the GCM to include collective momenta and/or the use of a more appropriate Gogny interaction have the potential to drastically improve future results. Continued development of the method presented in this work will also help resolve complicated questions regarding the mesh density, as well as self-interaction and self-pairing effects in the mean-field Hamiltonian. Finally, ongoing research regarding the microscopic treatment of spontaneous fission will maximise the usefulness of the GCM and TDGCM, produce new tools such as the projection techniques that have been developed, and further our understanding of nuclear physics as a whole.

\begin{acknowledgments}

The authors thank L. Robledo for the provision of \textsc{HFBaxial} and associated code, and thank L. Robledo and D. Regnier for valuable discussions. \\

The authors gratefully acknowledge the support of the CNRS/IN2P3 national supercomputing centre (Lyon, France), which provided computing and data processing resources needed for this project. This work also made use of the HPC resources of IDRIS and CINES under allocation No. 2024-AD010515531R1 made by GENCI. G.S. acknowledges support from the French National Research Agency (ANR) under the \textit{Chaire de Professeur Junior} program (Grant No. ANR-23-CPJ1-0154).
 
\end{acknowledgments}

\appendix

\begin{widetext}

\section{\label{sec:appendix_q20}The quadrupole moment operator in the \textsc{HFBaxial} harmonic oscillator basis}

\textsc{HFBaxial} utilises an axially symmetric harmonic oscillator basis with a maximum of $N$ principal shells, imposing the condition $2n_r + |m_z| + \frac{1}{q}n_z \leq N$, where the nonnegative integers $n_r$ and $n_z$ are the shell numbers of the two-dimensional ``radial'' (circular) HO in the $xy$ plane and the one-dimensional ``axial'' HO along the $z$ axis respectively, and the integer $m_z$ is the projection of the angular momentum of the radial HO on the $z$ axis. The radial and axial HOs have characteristic frequencies $\omega_r$ and $\omega_z$ respectively, which define the real axiality ratio $q = \omega_r / \omega_z$. Expression the quadrupole moment operator $\hat{Q}_{20} = C_{20}(\hat{z}^2 - (\hat{x}^2 + \hat{y}^2))$ into the basis of \textsc{HFBaxial} requires the transformation of the Cartesian position operators $\hat{x},\hat{y},\hat{z}$ into the new basis in a way that reproduces the correct quantum numbers $n_r,m_z,n_z$; this process is lengthy and not particularly enlightening. However, the result is a piecewise definition for the matrix $[Q_{20}]_{\alpha\alpha'}$, where the indices $\alpha$ and $\alpha'$ represent the relevant quantum numbers according to $\alpha \equiv (m_z,n_r,n_z)$:
\begin{align}
[Q_{20}]_{\alpha\alpha'} = \bra{\alpha\vphantom'} \hat{Q}_{20} \ket{\alpha'} = C_{20} \frac{\hbar}{m\omega_r} \delta_{m_z m'_z} \begin{dcases}
q(2n_z + 1) - (2n_r + |m_z| + 1) & \text{if } n'_r = n_r \text{ and } n'_z = n_z, \\
q\sqrt{n_z(n_z - 1)} & \text{if } n'_r = n_r \text{ and } n'_z = n_z - 2, \\
q\sqrt{(n_z + 2)(n_z + 1)} & \text{if } n'_r = n_r \text{ and } n'_z = n_z + 2, \\
\sqrt{n_r(n_r + |m_z|)} & \text{if } n'_r = n_r - 1 \text{ and } n'_z = n_z, \\
\sqrt{(n_r + 1)(n_r + |m_z| + 1)} & \text{if } n'_r = n_r + 1 \text{ and } n'_z = n_z, \\
0 & \text{otherwise.}
\end{dcases} \label{eq:q20_kernel}
\end{align}

\section{\label{sec:appendix_pfaffian}Pfaffian formalism for calculation of operator kernels}

The method of calculating an operator overlap kernel $O(q,q') = \bra{\phi(q\vphantom')} \hat{O} \ket{\phi(q')}$ using a Pfaffian was first proposed by Robledo in \cite{robledo2009,robledo2011} in order to resolve a phase ambiguity affecting the older Onishi formula \cite{robledo1994}. The formalism is explained in more detail by Bertsch and Robledo in \cite{bertsch2012}. The HFB theory is set out in the standard way, beginning with the definition of a single-particle basis defined by creation and annihilation operators $\chd_m, \ch_m$. Every state $\ket{\phi(q)}$ in the generator basis is a superposition of many-body Slater determinants written as
\begin{align}
\ket{\phi(q)} & = \prod_k \bh_k \ket{-} = \prod_k \left(\sum_m U_{mk} \chd_m + V_{mk} \ch_m\right) \ket{-}.
\end{align}
The Bogoliubov transformation associated with the quasiparticle operator $\bh_k$, is defined as usual by the matrices $U_{mk}$ and $V_{mk}$. Applying the standard Bloch-Messiah decomposition according to Ring \& Schuck \cite[\S7.2]{ringandschuck}, which expresses the canonical forms of $U$ and $V$ as $\bar{U} = D^\dagger U C^\dagger$ and $\bar{V} = D^T V C^\dagger$ respectively, the overlap kernel of an arbitrary one-body operator $\hat{O}$ is derived in Ref.~\cite{bertsch2012} as
\begin{align}
\bra{\phi(q)\vphantom'} \hat{O} \ket{\phi(q')} & = (-1)^{N/2} \frac{\mathrm{det} C^*\ \mathrm{det} C'}{\prod_{k=1}^{N/2} v_k v'_k} \mathrm{pf}\begin{bmatrix} V^T U & V^T \hat{O}^T {V'}^* \\ -{V'}^\dagger \hat{O} V & {U'}^\dagger {V'}^* \end{bmatrix}\!,
\end{align}
where the non-primed quantities $U,V,C,v_k$ are associated with $\ket{\phi(q)}$ and the primed quantities with $\ket{\phi(q')}$, and the size of the Bogoliubov basis is $N$ for both states (this is twice the number of Bogoliubov states obtained during HFB calculations, as both positive-parity states $\bh_k$ and negative-parity states $\bh_{\bar{k}}$ are included). The real scalars $v_k$ are the elements defining the $2 \times 2$ diagonal block matrices that form $\bar{V}$; they are obtained by diagonalising the density matrix $V^* V^T$, which has pairs of eigenvalues ${v_k}^2$. However, Carlsson and Rotureau \cite{carlsson2021} propose several simplifications of this formula which take advantage of the Bloch-Messiah decomposition. By applying certain factorisations presented in their letter and its accompanying supplementary material, the following expression for the kernel is obtained:
\begin{align}
\bra{\phi(q)\vphantom'} \hat{O} \ket{\phi(q')} & = \frac{(-1)^{n/2}}{\prod_{k=1}^{n/2} v_k \prod_{k=1}^{n'/2}v'_k} \mathrm{pf} \begin{bmatrix} [-\bar{V} \bar{U}]_{n\times n} & [-\bar{V} D^\dagger \hat{O}^T D' \bar{V}']_{n\times n'} \\ [\bar{V}' D^{\prime T} \hat{O} D^* \bar{V}]_{n'\times n} & [\bar{U}' \bar{V}']_{n'\times n'} \end{bmatrix}\!. \label{eq:kernel_pfaffian}
\end{align}
Assuming that $\ket{\phi(q)}$ and $\ket{\phi(q')}$ have $n$ and $n'$ non-empty quasiparticle states respectively allows the rows and columns of the submatrices corresponding to empty states to be removed (the subscript of each submatrix indicates its dimensions).

\end{widetext}

\bibliography{Projection_Paper_2026_NWTLau}

\end{document}